\documentclass[english,submission]{programming}

\usepackage[svgnames]{xcolor}
\usepackage[backend=biber]{biblatex}
\usepackage{multicol}
\usepackage{ccicons}

\lstdefinelanguage[programming]{TeX}[AlLaTeX]{TeX}{%
  deletetexcs={title,author,bibliography},%
  deletekeywords={tabular},
  morekeywords={abstract},%
  moretexcs={chapter},%
  moretexcs=[2]{title,author,subtitle,keywords,maketitle,titlerunning,authorinfo,affiliation,authorrunning,paperdetails,acks,email},
  moretexcs=[3]{addbibresource,printbibliography,bibliography},%
}%
\newcommand*{\CTAN}[1]{\href{http://ctan\textsc{.}org/tex-archive/#1}{\nolinkurl{CTAN:#1}}}

\paperdetails{
  perspective=engineering,
  area={Interpreters, virtual machines, and compilers, General-purpose programming, Data structures},
  license=cc-by,
}

\newcommand{\union}{\cup}
\newcommand{\intersection}{\cap}
\newcommand{\prodarrow}{\;\;\rightarrow\;\;}
\newcommand{\rsep}{\;\;|\;\;}

\definecolor{bgcolor}{HTML}{EFEFEF}

\newcommand{\lhfurl}{\url{https://aghorui.github.io/lhf/}}

\usepackage{appendix}
\usepackage{makecell}
\usepackage{hyperref}
\usepackage{listings}

\begin{document}

\title{LatticeHashForest: An Efficient Data Structure for Repetitive Data and Operations}
\subtitle{Achieving Efficiency through Deduplication and Immutability}

\author[a]{Anamitra Ghorui}[https://orcid.org/0009-0003-0707-2373]
\authorinfo{is a masters student at IIT Bombay.\\ Contact: \email{aghorui@cse.iitb.ac.in}}
\affiliation[a]{Indian Institute of Technology, Bombay}
\author{Uday P. Khedker}[https://orcid.org/0000-0002-5203-4494]
\authorinfo{
  is a professor at IIT Bombay working in the domain of programming languages
  and compilers. \\Contact:  \email{uday@cse.iitb.ac.in}.}
\affiliation{Indian Institute of Technology, Bombay}




\maketitle

\begin{abstract}





The domain of program analysis and optimization forms a very fundamental part of compilers and software toolchains. Analysis of entire programs as a single unit, or whole-program analysis, involves propagation of large amounts of information through the control flow of the program. This is especially true for pointer analysis, where, unless significant compromises are made in the precision of the analysis, there is a combinatorial blowup of information. This makes a precise analysis infeasible for large programs. Many attempts towards making pointer analysis scale have been made in the past. One of the key problems we observed in our own efforts is that a lot of duplicate data was being propagated, and many low-level data structure operations were repeated a large number of times.

We present what we consider to be a novel and generic data structure, \textit{LatticeHashForest} (LHF), to store and operate on such values in a manner that eliminates a majority of redundant computations and duplicate data in scenarios similar to those encountered in compilers and program optimization. LHF differs from similar work in this vein, such as hash-consing, ZDDs, and BDDs, by not only providing a way to efficiently operate on large, aggregate structures, but also modifying \textit{the elements} of such structures in a manner that they can be deduplicated immediately. LHF also provides a way to perform a \textit{nested} construction of elements such that they can be deduplicated at \textit{multiple levels}, cutting down the need for additional, nested computations.

We provide a detailed structural description, along with an abstract model of this data structure. An entire C++ implementation of LHF is provided as an artifact along with evaluations of LHF using examples and benchmark programs. We also supply API documentation and a user manual for users to make independent applications of LHF. Our main use case in the realm of pointer analysis shows memory usage reduction to an almost negligible fraction, and speedups beyond 4x for input sizes approaching 10 million when compared to other implementations.

Besides enabling more efficient and easier implementation of program analysis frameworks, we believe the problems LHF solves represent a particularly overlooked class of problems where an efficient, frequently accessed, in-memory deduplicating storage, and efficient element-level manipulation of aggregate data is the most useful. While we find the main application of LHF in the domain of compilers and optimization, we believe that the generalized idea of the mechanism, along with its peripheral features, can be used to effectively tackle a large number of problems outside the domain of compiler design. This paper attempts to describe this class of problems and draw similarities to it within and beyond compiler design.

\end{abstract}

\section{Introduction}
\label{sec:introduction}

This paper describes \textit{LatticeHashForest} (LHF), a data structure for fast, efficient element~level operations on and storage of aggregate data. LHF functions by assigning a unique (integer) identifier to each unique instance of aggregate data that is registered within it. An operation (say, set union) between any two or more elements is memoized with a hash table based-mapping from a tuple of unique identifiers representing the operands to a unique identifier representing the result. This ensures immutability and efficient deduplication when aggregate data is frequently accessed or modified. Experimental results show memory usage reduction to an almost negligible fraction, and speedups beyond 4x for inputs approaching 10 million against comparable implementations in all cases.

In our observations, LHF is most useful in situations where data is often duplicated and operations are often redundant, and where operations are often performed at the element level. This situation is frequently encountered in program analysis because large regions of a program have similar analysis data with minor differences. Hence, much of the context for LHF's creation and development lies within the domain of program analysis and data-flow analysis, which the later sections expand upon. In spite of this, the nature of how LHF is modeled and implemented by us expresses a generic data structure and can be potentially used outside the field of compiler design and optimization, likely in problems that have a resemblance to those found within this domain.

In the following sections, we give a background of how the general workings of LHF came to be conceived, an abstract model of LHF, and the underlying pattern of the problem that we have observed LHF to solve. We also provide an evaluation of LHF, both in terms of abstract, randomized benchmarks, and a real-world use-case of it within a pointer analysis implementation built by us. We also compare LHF against closely related data structures such as ZDDs, e-graphs, and disjoint sets (union-find).

Our hope is that LHF will find applications in a variety of domains, well beyond the applications that we show in \autoref{sec:applications} and possibly become a standard solution and tool for a potentially overlooked class of problems.

The concept of hash-consing is an integral part of LatticeHashForest. Earlier work in pointer analysis~\cite{barbar2021hash} has used hash-consing in a similar manner to how we have done here. There are also other mechanisms, such as zero-suppressed decision diagrams~\cite{wiki-zdd}, union find~\cite{wiki-unionfind}, and e-graphs~\cite{wiki-egraphs} that also attempt to tackle the problem of efficiently computing operations on aggregate values. After providing a description of LHF, in \autoref{sec:comparison}, we compare these mechanisms to the one that is described by this paper.

\section{Background}
\label{sec:background}

Program analysis and optimization are important parts of compiler design, program transformation, and vulnerability detection, and form the basis of building efficient programs from an abstract description provided by a user in the form of source code. Performing interprocedural analysis over the entirety of a program as a single unit is a challenge to do efficiently, as it requires taking into consideration all of the possibilities of the calling contexts of functions (context-sensitivity) as well as the control flow (flow-sensitivity) of each function~\cite{lfcpa}. This, if not handled properly, results in a combinatorial blow-up of information and thus makes the analysis infeasible for large programs.

This is especially true for data-flow analyses like pointer analysis, where we determine where a pointer points to within a program point. Information being propagated within the analysis may increase combinatorially due to transitive relationships between pointer variables. For an illustration of this phenomenon, consider \autoref{fig:ptarel}, which shows C language examples of pointer assignments that lead to vastly different possibilities of points-to relationships between variables.


From this point forward in this paper, our main focus in terms of program analysis shall lie within pointer analysis, as our work leading up to the conception of LHF has been an effort of optimizing our pointer analysis implementation.


\subsection{Brief Description of Pointer Analysis}

\begin{figure}[t]
	\centering
	\includegraphics[width=0.7\textwidth]{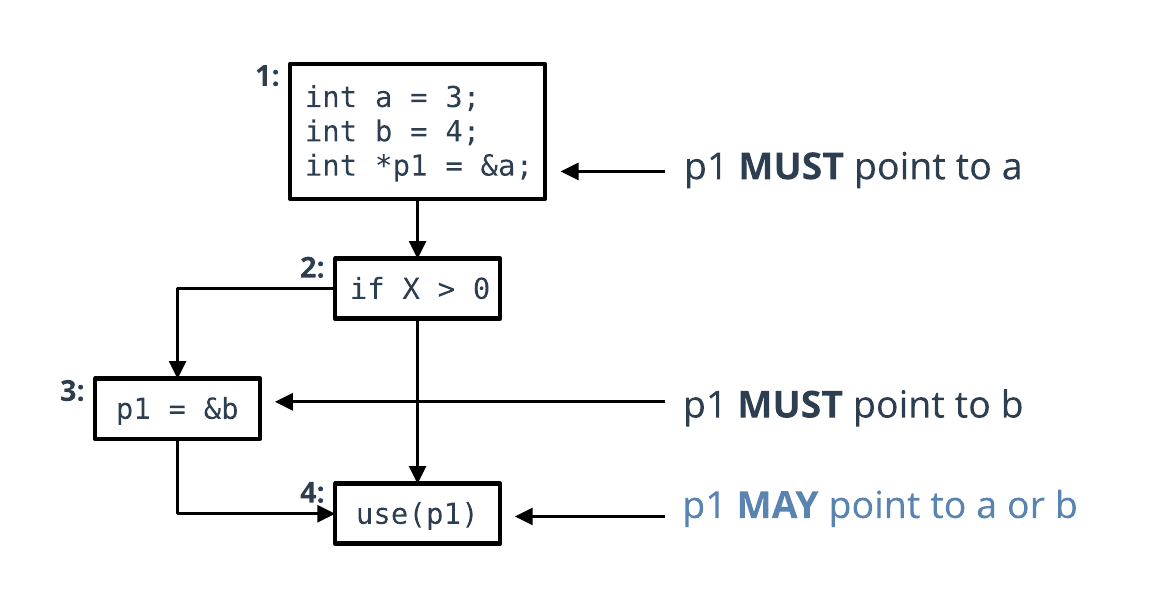}
	\caption{An example of flow-sensitive pointer analysis.}
	\label{fig:fispta}
\end{figure}

In this section, we provide a brief description of pointer analysis, specifically, points-to analysis\footnote{See \autoref{appendix:pta} for explanation on terminology.} and related concepts that is sufficient for understanding the content of the upcoming sections. Extra descriptions for pointer analysis techniques are provided in \autoref{appendix:pta}.


\paragraph{Representation of Relationships.}
Consider pointer variables $a$, $b$, and abstract locations $p$, $q$, $r$ within a program. If, at a program point, $a$ may point to all three locations (due to summarization across multiple paths), and $b$ must only point to $r$, we symbolically expose these facts as a mapping between pointers and sets of pointees. We call this a \textit{points-to set} and it is essentially the adjacency list representation of a graph.
\[
\lbrace 
	a \rightarrow \lbrace p, q, r \rbrace,
	b \rightarrow \lbrace r \rbrace
\rbrace
\]
\paragraph{Points-to Pairs.} An alternate way of representing this would be to create a set of points-to pairs. The above set of relationships would be exposed as:
\[
\lbrace 
	a \rightarrow p,
	a \rightarrow q,
	a \rightarrow r,
	b \rightarrow r
\rbrace
\]
The relevance of these two representations lies in the implementation of points-to analysis. The contrast between them is discussed further in \autoref{sec:initial-approaches}.

\paragraph{Running the Analysis.}
Points-to analysis is done through a process called fixed-point computation. To explain this intuitively, consider \autoref{fig:fispta} depicting a control flow graph of a program. In block 1, we can definitely deduce that \texttt{p1} points to the integer variable \texttt{a} in the data. Similarly, we can deduce that after block 3, \texttt{p1} will point to \texttt{b}. These facts `propagate' forward with the flow of control in the program, and as such, at the entry of block 4, due to the branch at the if statement at block 2, we end up with two possibilities of pointees for \texttt{p1}. We summarize the incoming data flow by taking the union of the two possibilities. Thus, the final representation at the exit of block 4 becomes:
\[
\lbrace 
	p1 \rightarrow \lbrace a, b \rbrace
\rbrace
\]
In the presence of recursion or loops, we have to keep repeatedly propagating, summarizing and updating values across control flow until the system reaches a 'fixed-point' and no further updates are possible. More explanation of the propagation of the semantics of points-to analysis is given in \autoref{appendix:pta}.

\paragraph{Precision, Flow-Sensitivity, and Context-Sensitivity.}

The above sample run of points-to analysis is a \textit{flow-sensitive} analysis. Most common pointer or alias analysis implementations do not take into consideration the control flow of the program, or the context of a call to a function in order to make them scale for large programs. This causes them to produce very imprecise results\footnote{See~\autoref{appendix:pta} for further discussion on scalability and precision.}. The pointer analysis implementation that we use attempts to preserve precision by maintaining these two factors. We discuss this in the next section.

\subsection{Liveness-Based Flow and Context-Sensitive Points-to Analysis (LFCPA)}

\label{sec:background:pta}
LFCPA~\cite{lfcpa} is an augmentation of normal flow and context-sensitive pointer analysis~\cite{ptatutorial} and is what we use in our C++ pointer analysis implementation. To enable interprocedural analysis, we further augment LFCPA with \textit{Value-Sensitive Contexts}, or VASCO~\cite{vasco}. Henceforth, we will refer to the combined mechanism as \textit{VASCO-LFCPA}. Here, we only provide brief explanations that are relevant to LHF.
\paragraph{General Idea of LFCPA.}
Liveness analysis~\cite{wiki-liveness} is another analysis technique by which we can determine what variables will be used later at the end of a program point. Unlike points-to analysis, this analysis is propagated in the reverse direction of the control flow. As an example of representation, if pointers $p1$ and $p3$ are live, we will represent them symbolically as:
\[
\lbrace p1, p3 \rbrace
\]
We use, in alternation, liveness analysis results of pointers in points-to analysis and vice versa to determine whether we should actually propagate points-to info or not, cutting down the information we actually need to propagate. Thus, instead of simply propagating points to information across the control flow, we also propagate variable liveness information. A sample representation at a program point may look like the following pair. The first set of the pair represents points-to information, whereas the second represents liveness information.
\[
(\lbrace 
	p1 \rightarrow \lbrace a, b \rbrace
\rbrace, \lbrace p1, p3 \rbrace)
\]
\paragraph{General Idea of VASCO.}
To enable interprocedural analysis and context-sensitivity, one of the classical methods is to differentiate contexts with call strings~\cite{callstrings}, which essentially captures the chain of calls that have been made up to the entry point of a function\footnote{There are special considerations made in the case of recursion.}. Call strings can turn out to be large and numerous, especially in large programs. Implementations might attempt to limit the size of the call string by using only the latest N elements of it, decreasing the precision of the contexts.

As an alternative, we may use value-sensitive contexts~\cite{vasco}, where we instead capture the incoming and outgoing data flow values at a given callsite and differentiate calling contexts on that basis. In the case of LFCPA, these would be the points-to and liveness information.

A point to note is that an analysis has to be \textit{re-run for each distinct context}, and therefore, a separate set of data-flow values is computed and stored for each context.

\section{Past Work and Motivations}
\label{sec:initial-approaches}
This section describes our past approaches to optimizing our pointer analysis implementation that led up to the motivations and conceptualization of LHF. This section also serves as a case study and justification of the utility of LHF.

\subsection{Scalability Issues}
Despite the augmentations made for VASCO-LFCPA being a net improvement over standard flow and context-sensitive pointer analysis, the issue of scalability in large programs persists. While attempts at improvement are still being made from the algorithmic perspective in our implementation, other avenues of improvement have been overlooked. It remained to be seen whether the lack of scalability of our VASCO-LFCPA implementation is due to the algorithm itself or the implementation.

The data structures used to run the algorithm were naive in nature and inefficient, essentially amounting to keeping a C++ STL\footnote{``Standard Template Library''} \texttt{std::set} object for liveness info, and a \texttt{std::map} of \texttt{std::set}s for points-to info. The size of the data quickly adds up when compounded with the fact that it effectively gets multiplied by the number of contexts generated.

Performing operations on data-flow values, like union and equality-checking, were implemented in an ad-hoc manner with either $O(n)$ or even $O(n^2)$ functions. This is despite the fact that \texttt{std::map} and \texttt{std::set} are ordered containers. The implementation simply used pointers that point to the value/variable representation record\footnote{This is an abstraction over \texttt{llvm::Value} in our system.} (henceforth referred to by the blanket term `operand'), and no criterion for ordering was defined on them. Hence, there was a very visible need for creating an alternate solution for representing data and operations on data in our implementation.

\subsection{Observations from Past Experience}
\label{subsec:initial-approaches:past-exp}

Informal observations of our past analysis work and others~\cite{barbar2021hash} have yielded the following three general trends for a points-to analysis:

\begin{description}
\item[Sparseness.] Points-to information is usually very sparse in programs.
\item[Low Information Entropy.] A high number of duplicates exist in the information across the control flow.
\item[Computational Redundancy.] Much of the computations performed in an analysis are redundantly performed on the same data.
\end{description}

Besides improving our data representation, these observations led us to believe that it would make sense to create a \textit{deduplication and storage engine} that would allow us to get rid of all data redundancies within our analysis.

\subsection{The Deduplication Mechanism}
\label{subsec:initial-approaches:dedup}
The deduplication mechanism described below forms a core part of LHF, as well as the initial approaches we used. It maps a unique object to a unique integer identifier. When an object is queried in the mechanism, it does the following:

\begin{enumerate}
\item If the object is already mapped to an identifier, return that identifier.
\item Otherwise, store the object, map the object to a new identifier, and return that identifier.
\end{enumerate}

When an \textit{identifier} is queried in the mechanism instead, a read-only reference to the object is returned. A key point to note here is that the objects stored (or registered) in this mechanism are kept \textit{immutable}, which is important for maintaining uniqueness and correspondence of the identifiers. An implementation of this mechanism used in our work is provided in \autoref{appendix:deduplication}, but in essence, it uses a hash table-based map and a storage vector with pointers to the location of the data. Any future references to deduplication in this document will specifically refer to this mechanism.

\subsection{First Idea of the Data Structure}
Based on these observations, we drafted a system wherein we would, through the deduplication mechanism:
\begin{enumerate}
\item Assign a unique integer to each operand/variable, and use a pair of these integers as a points-to pair.
\item Assign a unique integer to each points-to pair, and put these integers into a sorted vector to use as our points-to set. Use the integer identifiers for individual operands in a sorted vector as our live variable set.
\item Assign unique integers to the points-to sets and live variable sets, and use them as our points-to and liveness representation in the analysis respectively.
\end{enumerate}
This approach allows us several advantages despite introducing additional levels of indirection for access. The first being that attribute-based comparisons of individual operands are ruled out. The second being that comparisons between points-to pairs are reduced to a single integer. The third is that, by virtue of being a sorted list, the majority of aggregate operations on the sets are $O(n)$ in the worst case.


\section{LatticeHashForest (LHF)}
\label{sec:latticehashforest}

While our initial approach let us reduce the effective amount of redundancy in the system, it still did not solve the problem of reducing the amount of redundant operations that are performed. Another big problem was that the points-to pair representation was not compatible with the pointer-pointee set representation that VASCO-LFCPA used, and introduction would have essentially amounted to a full rewrite.

Several ideas were floated around, such as performing these operations lazily by building `composition' trees\footnote{This approach is similar to what is used in ZDDs. We comment on this in \autoref{sec:comparison}.}, and then only calculating them when needed. However, we had concerns in regards to the storage, traversal and flattening of the tree. We also did not know how big these trees would become or if we would be able to still provide a unique identifier for each points-to set without flattening them.

\subsection{Deeper Implications}
\label{subsec:latticehashforest:deeper_implications}

\begin{figure}[t]
	\centering
	\includegraphics[width=0.65\textwidth]{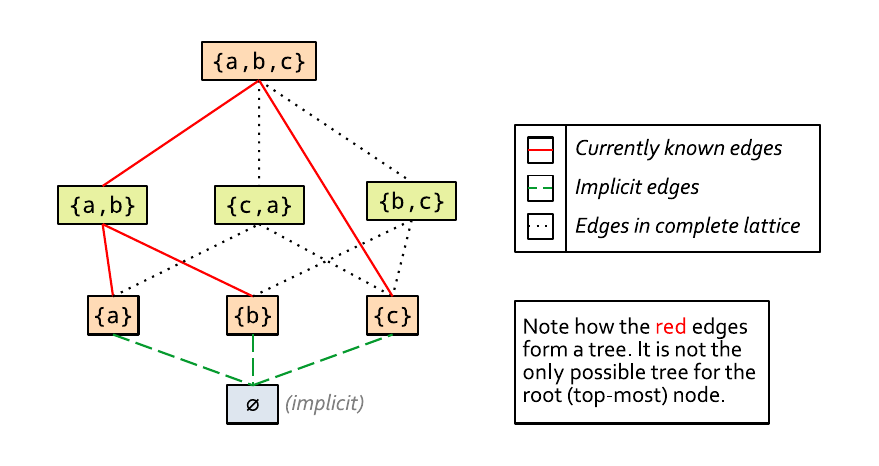}
	\caption{Demonstration of incremental lattice building in LHF.}
	\label{fig:lhfabstract}
\end{figure}

Around this time, while searching for similar data structures, we came across a paper~\cite{barbar2021hash} that had a very similar idea to us, but they also had implemented a method for removing operational redundancy in points-to sets\footnote{\label{footnote:pts} It is important that we point out that what the paper~\cite{barbar2021hash} terms as a `points-to set' is in fact what we refer to as a pointee set in this paper, that is, not a collection of mappings from a pointer to a set of pointees, but only a set of pointees.}. They keep a hash table mapping sets to unique integers as we have, but alongside that, they also keep hash table mappings from a tuple of two identifiers to another identifier for representing memoized binary operations.

\paragraph{Incremental Lattices.} A conclusion that the authors do not make is that the act of memoizing operations, say, for unions of sets, in this manner essentially creates edges, starting from the operands to the result, thus implicitly creating a composition tree. This composition tree is not necessarily unique, as there may be many sets for which the union may give the same result, and similarly, the operands themselves may have multiple compositions. These non-unique composition trees, when merged, form a \textit{forest}, essentially reducing to an incremental construction of a \textit{lattice over the subset relation}, with the join operation being union.

In \autoref{fig:lhfabstract}, we see an illustration of this fact with the set $\{a,b,c\}$. This set can be incrementally created by taking the union of $\{a\}$ and $\{b\}$, then taking the union of $\{a, b\}$ and $\{c\}$, or by taking a union between $\{b, c\}$ and $\{a\}$, and so on. All of these results are memoized by the \textit{hash table}, which allows it to implicitly and incrementally represent the edges of multiple composition trees at once, and by extension, an entire lattice. 

\paragraph{Analogy with Virtual Memory.} Another conclusion we draw from this is that representation of values by unique identifiers, in a way, allows us to abstract away data itself rather than the location where the data is written. That is, the `pointers' in this case point to unique data rather than unique memory locations. This opens up opportunities similar to what operating systems are able to do with virtualizing the memory an application accesses in a user-invisible manner. We discuss a few possibilities in \autoref{sec:other-features} and \autoref{appendix:proposed-features}.

\paragraph{Generalization.} The general structure that is exposed here can possibly be used not just for points-to analysis, but for representing any form of abstract data and operations on that data. Much of the upcoming text is concerned with exploring this line of thought. 

From these observations, we derive the namesake of \textit{LatticeHashForest}, and the final idea of the data structure.

\subsection{Definitions and Implementation}
\label{subsec:latticehashforest:implementation}

\begin{figure}[t]
	\centering
	\includegraphics[width=0.8\textwidth]{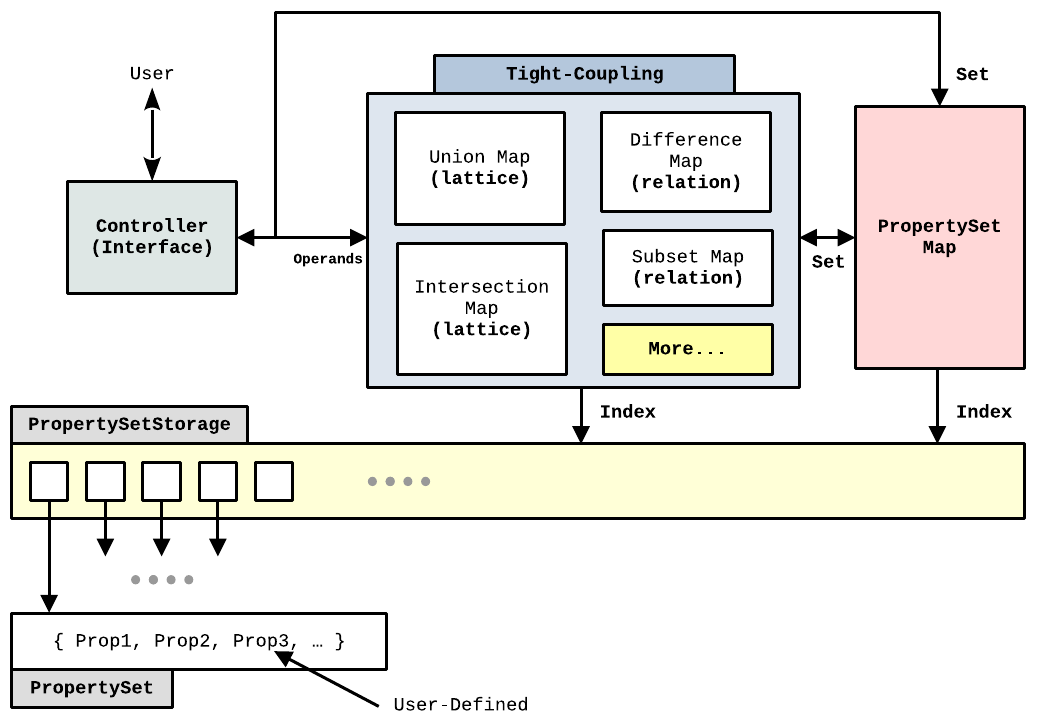}
	\caption{General structure of LatticeHashForest.}
	\label{fig:lhf}
\end{figure}

The general structure of LHF is as depicted in \autoref{fig:lhf}. For this, we define the term \textit{Property} to refer to the atomic data that can be stored by LHF, for example, points-to pairs and live variables. We use \textit{property set} to refer to the sets of these properties. The integer identifiers assigned to these are called \textit{indices}. We implement a deduplication engine in a similar way to that described in previous sections, visible in the figure as \texttt{PropertySetMap} and \texttt{PropertySetStorage}. Finally, we add a series of hash maps (called \textit{operation maps}) for memoizing operations.

Besides these, we have another mapping known as the \textit{subset map}, which stores subset relationships inferred from other operations. This allows us to trivially deduce some operations. This `information sharing' between operations is why we refer to these maps as being \textit{tightly coupled}.

We refer to property sets as being \textit{bound} or \textit{closed} under the operations that LHF provides. The LHF both defines the type of data a property set can hold as well as the operations on it. In some sense, it forms an algebraic structure.

\paragraph{Implementation.} The entire implementation\footnote{\lhfurl} is written as a C++ header-library. It makes heavy use of template meta-programming to enable usage with any data type, as long as the data type is less-than comparable, equality-comparable and hashable, that is, the data type is \textit{totally ordered}. Enforcing partial ordering would have been enough, but having an explicit equality operation allows for a more efficient implementation. The default constituent data structures are all from the C++ STL, but can be easily replaced with other implementations. 

\subsection{Sample Demonstration}
\label{sec:latticehashforest:sampledemonstration}

\begin{figure}[t]
	\centering
	\includegraphics[width=0.72\textwidth]{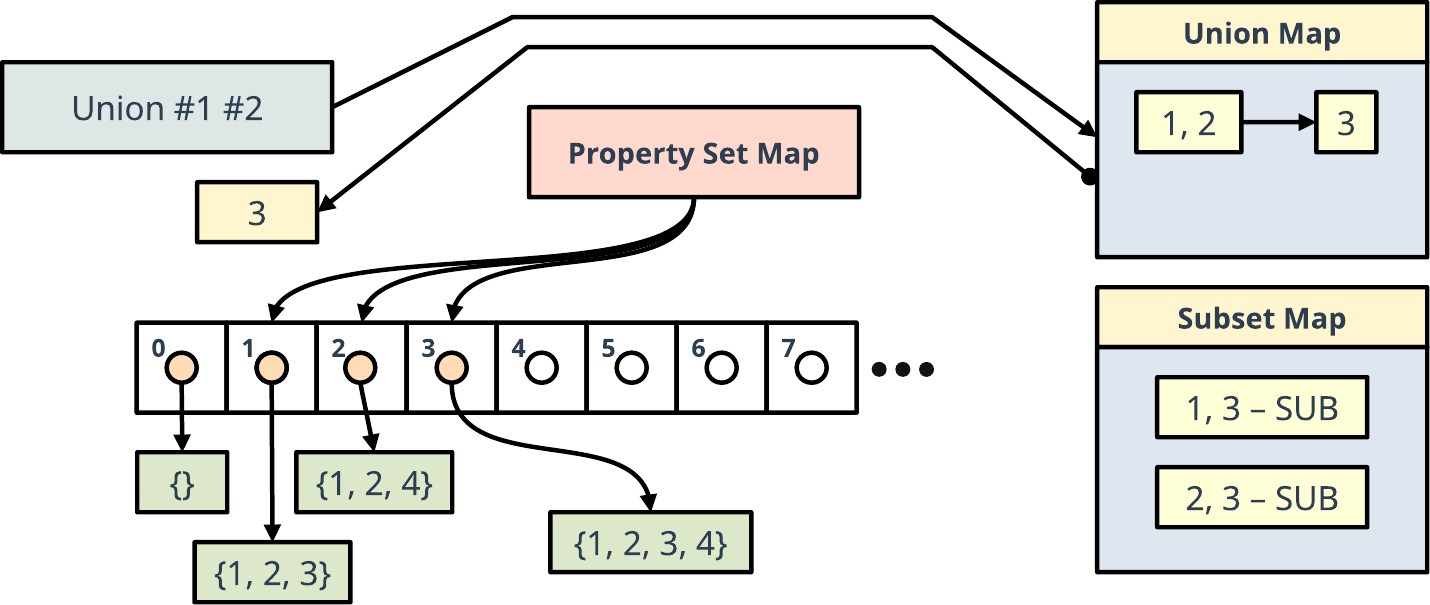}
	\caption{The final configuration of LHF based on the steps provided in \autoref{sec:latticehashforest:sampledemonstration}.}
	\label{fig:lhfconfig}
\end{figure}

Let us consider an `empty' LHF that is configured to store and operate on sets of integers, and we want to perform the union of \texttt{\{1, 2, 3\}} and \texttt{\{1, 2, 4\}}.
\begin{enumerate}
\item First, we \textit{register} \texttt{\{1, 2, 3\}} and then \texttt{\{1, 2, 4\}} into it. This gives us the identifiers \texttt{1} and \texttt{2}, respectively, by the deduplication process in \texttt{PropertySetMap}.
\item Next, to actually perform the union, we perform a call to the union operation with these two identifiers.
	\begin{enumerate}
	\item First, LHF checks whether either of the operands is \texttt{0}, which is a special index reserved for the empty set.
	\item Seeing this isn't the case, LHF tries to look up the tuple \texttt{(1, 2)} in the union operation map, but it does not find it.
	\item After this, it looks in the subset map if the operation can be trivially deduced with a subset relation, but it fails.
	\item  It determines that it has to actually perform the operation, which it does and creates the new set \texttt{\{1, 2, 3, 4\}}. This gets registered in the LHF with the index \texttt{3}, and this is the result that gets returned to the user.
	\end{enumerate}
\item  LHF records this new union that has been performed, as well as the subset relations inferred from the operation. Any future requests for the same or adjacent operations will result in the memoized index being returned.
\end{enumerate}
The final state of the sytem is shown in \autoref{fig:lhfconfig}. We provide a sample C++ program for this demonstration in \autoref{appendix:sample-code:simple}.

%

\subsection{Nesting LHFs and Points-to Representation}
\label{subsec:abstract-model:nesting-mechanism}

\begin{figure}[t]
	\centering
	\includegraphics[width=0.7\textwidth]{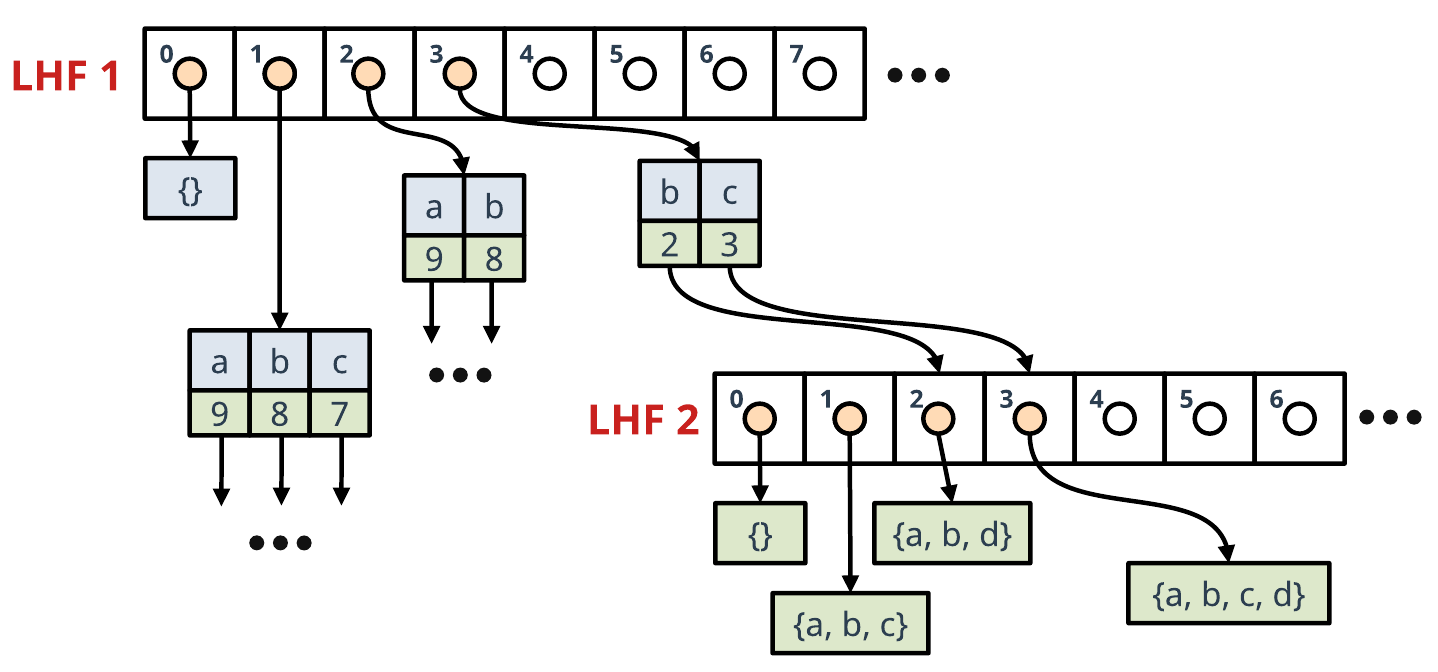}
	\caption{An illustration of nesting in points-to sets.}
	\label{fig:nesting2}
\end{figure}

While the base structure of LHF is useful for exploiting data redundancy on sets of elements, it is not by itself equipped for dealing with \textit{multi-faceted} redundancy. Consider the following points-to set (with integers) as an example.
\[
	\{ 
		2 \rightarrow \{ 3, 4, 5 \},
		3 \rightarrow \{ 3, 4, 5 \},
		4 \rightarrow \{ 4, 5, 6 \}
	\}
\]
If we were to decompose this into a set of points-to pairs, instead of three elements, we would have nine instead. If we were to use the pointer and set of pointees directly as the \texttt{property} in LHF, it still does not capture the fact that $2$ and $3$ have the same pointee set, and increases the cost of hashing and comparisons per-element to $O(n)$ (where $n$ is the number of pointees).

To capture the benefits of LHF for pointee sets, we \textit{nest} the two LHFs such that the points-to sets become a list of key-value pairs, wherein the key is the pointer, and the value is an \textit{index} within the pointee set LHF. For example, if the indices for \texttt{\{3, 4, 5\}} and \texttt{\{4, 5, 6\}} are 1 and 2, respectively, the above set becomes:
\[
\{ 
	(2, 1),
	(3, 1),
	(4, 2)
\}
\]
We refer to these augmented elements as \textit{property elements}. LHF can be configured to have any number of nested LHFs as a property element tuple, and the nested LHFs \textit{themselves} can also nest LHFs. We call this `network' a \textit{construction of LHFs}. This can possibly be used to represent any arbitrary data structure. \autoref{fig:nesting2} gives an illustration of another case of nesting.

\paragraph{`Natural' Nesting.} Operations in this nested LHF have to be implemented in a fashion that conveys the same semantic information as our previous representations. To do this, we implement a general algorithm as follows for a binary operation $A * B$:

\begin{enumerate}
\item Find the common elements in A and B based on the \textit{key} of each property element. Let this be $K = A \intersection B$.
\item Remove these elements from A, and perform the operation with the empty set on the \textit{right} of the operator:
$(A - K) * \emptyset$. Add it to the result set. This makes the operation `behave' as if the other set did not contain the common elements. We demonstrate the intuition in the upcoming example.
\item Remove these elements from A, and perform the operation with the empty set on the \textit{left} of the operator:
$\emptyset * (B - K)$. Add it to the result set.
\item Now, for each key $k$ in $K$, perform the operation on the corresponding values $v_1$ (from $A$) and $v_2$ (from $B$), creating the value $(k, v_1 * v_2)$. Add all of these pairs to get the final resultant set.
\end{enumerate}

This algorithm\footnote{The C++ implementation uses a specialization of this algorithm. Please see~\autoref{appendix:lhf-implementation-details}.} is what we refer to as the \textit{natural nesting behavior}, and is what LHF does by default in the case of a nested construction. This can be used for any binary set operation, commutative or non-commutative, and the behavior can be changed if required by providing a custom definition of it. Here is an example of performing a union of two points-to sets with the nested mechanism and the example shown earlier:
\[
	\{ (2, 1), (3, 1), (4, 2) \} \union \{ (4, 1), (5, 1) \}	
\]
$\{(2,1), (3,1)\}$ are the non-common elements in the first set. Performing a union with $\emptyset$ results in the same set. Similarly, $\{(5, 1)\}$ is the uncommon element from the second set. Finally, we take the common elements and perform the union recursively, giving us the set $\{(4, 3)\}$, where $3$ is the index of $1 \union 2$. Hence, the final set is:
\[
	\{ (2, 1), (3, 1), (4, 3), (5, 1) \}	
\]
It is important to note that the semantics of union or any other operation $*$ on a property set is dictated by the LHF it is closed under. Sample LHF code for nesting is provided in \autoref{appendix:sample-code:points-to} and \autoref{appendix:sample-code:points-to-2elem}.

\subsection{Usage of Nesting in VASCO-LFCPA}
\label{subsec:latticehashforest:lfcpa-nesting}
VASCO-LFCPA, as we mentioned earlier, requires a tuple of two values: a points-to set, and a set of live variables. For points-to data, we implement this with a nested LHF, similarly to what has been described in the previous section. For live variables, note that the data we use to represent pointee sets, which is a set of variables/operands, is \textit{exactly the same} as the data we use to represent live variables. This means that we can \textit{reuse} the LHF used for pointee sets for live variables and receive the benefit of memoization from both usages. We provide a model for seeking such optimizations in~\autoref{subsec:abstract-model:lhf-const} and provide more discussion in~\autoref{appendix:cfgsignificance}.

\section{An Abstract Model of LHF}
\label{sec:abstract-model}
In this section, we attempt to provide a formal modelling of LHF. In essence, LHF defines operations on \textit{sets of pairs of a key and a value representing a tuple of multiple sets}. Each property set of LHF forms the following general structure:
\[
\label{eqn:pset}
\Theta = L_k:\lbrace (\Phi_1, \Psi_1), (\Phi_2, \Psi_2), (\Phi_3, \Psi_3), ... \rbrace
\]
Where $L_k$ is the LHF under which the property set is \textit{closed}, $\Phi_n$ is some scalar object which we refer to as the \textit{key}, and $\Psi_n$ is a tuple of sets of the form $( \Theta_1, \Theta_2, \Theta_3, ..., \Theta_n)$ which we refer to as the \textit{value}.  This essentially forms a tree of recursive sets, which we refer to interchangeably as \textit{`nested sets'}. To form the base case of the recursion, i.e., a \textit{non-nested set}, we define a special symbol substituted for $\Psi$ called $\lambda$. In other words, it forms a leaf node of the tree:
\[
\Lambda = L_k:\lbrace (\Phi_1, \lambda), (\Phi_2, \lambda), (\Phi_3, \lambda), ... \rbrace
\]
We define a primitive operation $\alpha$ that extracts the keys from the key-value pairs of the property sets:
\[
\alpha(L_k:
	\lbrace 
		(\Phi_1, \Psi_1), 
		(\Phi_2, \Psi_2), 
		(\Phi_3, \Psi_3), ...\rbrace) =
L_k: \lbrace \Phi_1, \Phi_2, \Phi_3, ...\rbrace
\]
\paragraph{Operations on Property Sets.}

The operations in an LHF are defined on $\alpha(\Theta)$, that is, the keys of a property set. This allows us to enable recursive operational behavior on $\Theta$ in a manner such that we can use a common identifier for an operation across all LHFs, but the semantics of an operation are left to the individual LHFs. For example, if we have the operation $\union$ in an LHF containing sets of integers mapping to sets of strings, we can have $\union$ implement a concatenation operation in the LHF of strings. Operations can be variadic in nature and are not limited to one or two operands.

%

\subsection{Modelling Constructions of LHF}
\label{subsec:abstract-model:lhf-const}
We may symbolically represent constructions of LHF in the following manner:
\[
\text{LHF}(\textit{Prop}, \textit{OP}, \textit{Nest}_1, \textit{Nest}_2 ...)$$
\]
%
Where $\textit{Prop}$ is the set of property sets, $OP$ is the set of operations closed on $\{\alpha(\Theta) | \Theta \in \textit{Prop}\}$, and $\textit{Nest}_i$ is another LHF definition. In fact, this representation allows us to form an entire context-free grammar as depicted in \autoref{fig:lhf-cfgrammar}. We discuss the significance of this in \autoref{appendix:cfgsignificance}. As an example, the points-to set depicted in the previous section would be expressed as follows.
\[
\begin{aligned}
\textit{PointeeSet} &= \text{LHF}(\texttt{int},\{ \union, \intersection, ... \}) \\
\textit{PointsToSet} &= \text{LHF}(\texttt{int},\{ \union, \intersection, ... \},\textit{PointeeSet})
\end{aligned}
\]
\begin{figure}[t]
\fcolorbox{bgcolor}{bgcolor}{\begin{minipage}{\textwidth}
\[
\begin{aligned}
\nonumber\textbf{LHFConstr} &\prodarrow \text{LHF}(\textbf{Prop}, \textbf{OP}, \textbf{Nesting}) \rsep \text{LHF}(\textbf{Prop}, \textbf{OP})\\
\nonumber\textbf{Nesting} &\prodarrow \textbf{LHFConstr}, \textbf{Nesting} \rsep \textbf{ LHFConstr} \\
\nonumber\textbf{Prop} &\prodarrow \text{<identifier>} \rsep \text{<set of property sets>} \\
\nonumber\textbf{OP} &\prodarrow \text{<set of operations>}
\end{aligned}
\]
\vspace{1pt}
\end{minipage}}
\caption{Illustration of a context-free grammar for LHF constructions.}
\label{fig:lhf-cfgrammar}
\end{figure}

\vspace{-15pt}
\subsection{Defining The General Recursive Operation of a Property Set}
\label{sec:abstract-model:operations}
We first start by defining the following primitive operations:

\begin{description}
\item[$\text{Restrict}(\Theta', \theta)$] allows us to create a set of tuples given a set of keys, e.g., $\alpha(\Theta)$ by restricting its domain with a property set parameter:
\[
\text{Restrict}(\Theta', \theta) = \lbrace (\Phi_1, \Psi'_1), (\Phi_2, \Psi'_2), (\Phi_3, \Psi'_3), ...) \rbrace \\
\]
\item[$\text{Restrict}(\Lambda, \theta)$] reconstructs a non-nested property set:
\[
\text{Restrict}(\Lambda, \theta) = \lbrace (\Phi_1, \lambda), (\Phi_2, \lambda), (\Phi_3, \lambda), ...) \rbrace
\]
\item[$\text{Common}(\Theta_1, \Theta_2)$] retrieves the keys that are common in $\Theta_1$ and $\Theta_2$:
\[
\text{Common}(\Theta_1, \Theta_2) = \alpha(\Theta_1) \intersection \alpha(\Theta_2)
\]
\item[$\text{From}(\Theta)$] returns the corresponding LHF of $\Theta$
\[
\text{From}(L_k:\lbrace ... \rbrace) = L_k
\]
\item[$\text{Value}(\Theta)$] returns the child property sets from $\Theta$ for some key $\Phi$
\[
\text{Value}(\Theta = \{(\Phi, \Psi), ...\}, \Phi) = \Psi
\]
\item[$\text{Nested}(L)$] returns whether $L$ is a nested LHF or not.
\item[$\text{Operate}(L, OP, \theta_1, \theta_2, ...)$] performs the operation $OP$ present in LHF $L$ on keys $\theta_1, \theta_2, ...$, where $\theta_n = \alpha(\Theta_n)$.
\end{description}

\vspace{1em}

\noindent Now, we define the operation in for a non-nested property set ($\Psi = \lambda$): 
\[
\text{NormalOperation}(L, OP, \Theta_1, \Theta_2, ...) = \text{Restrict}(\Lambda, \text{Operate}(L, OP, \alpha(\Theta_1), \alpha(\Theta_2), ...))
\]
$\textit{NormalOperation}$ simply performs the n-ary operation $OP$ on the keys of each set $\Theta_i$, and then applies the $\textit{Restrict}$ primitive to turn it back into a non-nested property set.

Next, we define nested operations based on natural nesting on sets $\Theta_1$ and $\Theta_2$. We begin by defining a few symbols for our convenience:
\[
\begin{aligned}
\theta_1 &= \alpha(\Theta_1) \\
\theta_2 &= \alpha(\Theta_2) \\
\Pi &= \text{Common}(\theta_1, \theta_2) \\
\end{aligned}
\]
$\textit{NestedOperation}$ is responsible for the natural nesting logic as described in \autoref{subsec:abstract-model:nesting-mechanism}, and gives the result of the nested computation as output. It is composed from a union of 3 sets. The first one is with the non-common elements in $\Theta_1$, the second one for non-common elements of $\Theta_2$, and finally, we use $\textit{Recurse}$ to recursively compute the result for the common elements.
\[
\begin{aligned}
\text{NestedOperation}&(L, OP, \Theta_1, \Theta_2) 
= \\
&\;\text{Restrict}\!\left(\Theta_1, \text{Operate}(L, OP, \theta_1 - \Pi, \emptyset)\right) \\
&\union\; \text{Restrict}\!\left(\Theta_2, \text{Operate}(L, OP, \emptyset, \theta_2 - \Pi)\right) \\
&\union\; 
	\left\{ \, (\Phi, \text{Recurse}(\Psi_1, \Psi_2)) \;\middle|\;
		(\Phi, \Psi_1) \in \Theta_1 ,\; (\Phi, \Psi_2) \in \Theta_2 \, \right\}
\end{aligned}
\]

\[
\begin{aligned}
\text{Recurse}(\Psi_1, \Psi_2) =&
\left(
\text{Operation}\!\left(CL, OP, \Theta_i', \Theta_i''\right) 
\right)_{i=1}^n \\
\vspace{0.5pt}
&\begin{aligned}
\textit{Where}\;\;\;& CL = \text{From}(\Theta_1'), \\
&\text{From}(\Theta_i') = \text{From}(\Theta_i''), \\
&\Theta_i' \in \Psi_1, \Theta_i'' \in \Psi_2, n = |\Psi_1| = |\Psi_2| \\
\end{aligned}
\end{aligned}
\]
\textit{Operation} is the driver function for actually performing an operation between two property sets in an LHF. As can be seen above, \textit{Recurse} is defined in terms of \textit{Operation}.
\[
\begin{aligned}
\text{Operation}(L, OP, \Theta_1, \Theta_2) =
\begin{cases}
  \text{NestedOperation}(L, OP, \Theta_1, \Theta_2)
    & \text{if } \text{Nested}(L), \\[6pt]
  \text{NormalOperation}(L, OP, \Theta_1, \Theta_2)
    & \text{otherwise.}
\end{cases}
\end{aligned}
\]
It is worth pointing out that the nested operation is only implemented for two operands here. The reason for this is that modeling associativity  in a generic manner for any operation would make it very complicated. This behavior can be changed and augmented for the individual LHF, which would require \textit{NestedOperation} to be redefined. In such cases, sufficient indication or explanation should be present for the augmented LHF.

\subsection{Modelling the Abstract Problem that LHF Solves}
\label{subsec:abstract-model:problem}

Registering data and memoizing operations in the manner we do in LHF may raise the question of how it may manage increasing memory allocations and possible blow-up. In all of our empirical observations on integrating LHF into VASCO-LFCPA, we use less memory with LHF than a naive implementation for the same purpose. The reasons for this are partly the ones we describe in \autoref{subsec:initial-approaches:past-exp}, but that does not necessarily mean that given an infinite amount of time, LHF will not take an infinite amount of memory as the system receives data.

\paragraph{Closed-World Computation.} The problem that we believe LHF solves most effectively, is in use cases where \textit{no new data besides what is originally available} is ever added to the system. All data is \textit{synthesized} from this original data, which we choose to call the \textit{corpus}. The corpus can be imagined as a set of cells that are initially filled with some data. The number of cells can increase by some heuristic of the system, but the new cells can only contain the original or synthesized data. We term this problem a \textit{`closed-world computation'}. We believe that the finite size of the original data in the system places a strict bound on the amount of unique data that can be generated, which works to our advantage.

\paragraph{Relation With VASCO-LFCPA.} This is in direct correspondence with how VASCO-LFCPA performs analysis. The only original data available in the system is what is found in the input program. We compute and propagate data-flow across instructions and functions, and the creation of new contexts based on VASCO's heuristics creates an entirely new instance of computed data with different initial conditions.

We do not have a formal proof for these claims. However, in \autoref{sec:evaluation}, we provide empirical results of an implementation of this model performing better in the case of LHF instead of an equivalent naive implementation. We also provide empirical results for the opposite case, where LHF does \textit{not} perform well against a naive implementation.

\section{Comparison With Similar Data Structures}
\label{sec:comparison}

The activities that LHF performs for processing operations may not be very different from what may be seen in many other contexts, such as database applications, caching mechanisms, and other data structures. This section provides comparisons with similar data structures that we believe are similar to LHF in terms of their application domain.

\paragraph{BDDs and ZDDs.}
\label{subsec:comparison:bdds-zdds}
Binary Decision Diagrams (BDDs) and Zero-suppressed Decision Diagrams (ZDDs)~\cite{taocp4A,wiki-zdd} are data structures that can be used to give compact representations of boolean functions and sets. ZDDs in particular are used for compact set operation representations in the form of acyclic graphs. This however makes concretizing them to a `flat' set on which arbitrary operations can be performed a costly task. A binary encoding is also enforced for all set elements, unlike LHF.

\paragraph{Union-Find/Disjoint Set Data Structures.}
\label{subsec:comparison:union-find}
Union-Find~\cite{wiki-unionfind} structures represent unions of disjoint sets in the form of trees, and allow us to quickly determine membership of elements from the representation. This, however, also does not allow us to flatten and perform arbitrary operations on the data easily. Performing operations other than union is also, at the very least, not a trivial task.

\paragraph{E-Graphs.}
\label{subsec:comparison:egraphs}
An expression graph (e-graph)~\cite{wiki-egraphs} is a data structure very similar in nature to LHF in the sense that it uses hash-consing~\cite{wiki-hashcons} as part of its mechanism. It builds equivalence sets of expressions in a language and enables querying equivalence through building a union-find structure. Hash-consing allows them to compose sub-expressions and assign an equivalence class ID to the composition.

\paragraph{Bit-Vectors and Sparse Bit-Vectors.}
\label{subsec:comparison:sparse-bit-vectors}
Bit-vectors enable efficient set operations with suitable binary encodings, but their utility in points-to analysis is limited by the unbounded growth of elements at runtime. Sparse bit-vectors remain a possible future direction, though their effectiveness in LHF’s multi-level memoization is uncertain.

\paragraph{Hash Consed Points-to Sets.}
\label{subsec:comparison:hash-consing}
We provide a few more notes on the paper we mentioned earlier in \autoref{subsec:latticehashforest:deeper_implications}~\cite{barbar2021hash}. The points-to value representation~\footnote{Please see \autoref{footnote:pts}.} they use is not a nested structure in the sense that we have described in this paper. They only perform operation memoization on the pointee sets. This means that storage of points-to sets as a whole incurs a higher cost.

\paragraph{Summary.}
\label{subsec:comparison:summary}
The conventional solutions available to date seem to make heavy use of graph and tree representations, and bitvector-like structures of data to cheapen operations  on sets. None of the solutions we explored allow a way to quickly perform arbitrary reads and operations on sets or map-like data structures. We put forward this as one of the pieces of evidence of LHF being a potentially novel data structure.

\section{Other Features}
\label{sec:other-features}
Some nascent features we are exploring in regards to LHF are described below.
\paragraph{Parallelization.}
\label{subsec:other-features:parallelization}
Operations over an LHF can be naively parallelized through read-write mutexes on the data components of LHF. Immutability of property sets works to our favor here. This possibly opens the door to solutions like worker threads that can optimistically perform future expected operations to speed up the actual computation.
\paragraph{Memory Management.}
\label{subsec:other-features:memory-management}
While we consider LHF to have good memory efficiency when the abstract problem described in~\autoref{subsec:abstract-model:problem} holds, we want to see whether this scope can be expanded on. Currently, LHF implements a way to manually evict sets. For this, the user must ensure that no references to the set being evicted are present in the program. If a queried operation results in the ID of the evicted set, the operation is recomputed and brought back into LHF's storage. This may, however, conflict with the immutability guarantees in LHF. This direction still needs to be explored. Besides this, we have a few more proposed features detailed in \autoref{appendix:proposed-features}.

\section{Evaluation}
\label{sec:evaluation}

Here, we provide an evaluation based on the following claims made in the paper:

\begin{description}
\item[Claim 1:] 
	LHF performs \textit{worse} than an equivalent naive approach in cases where the problem is not a closed-world computation, and/or does not meet the criteria in \autoref{subsec:initial-approaches:past-exp}.
\item[Claim 2:]
	LHF performs \textit{better} than an equivalent naive approach in cases where the problem is a closed-world computation, and/or meets the criteria in \autoref{subsec:initial-approaches:past-exp}.
\item[Claim 3:]
	LHF performs \textit{better} than the original naive approach in VASCO-LFCPA.
\end{description}

Evaluation materials and source code for all three of these claims are provided as part of the artifact, except for the SPEC CPU 2006  Benchmarks~\cite{speccpu2006}, due to them being licensed software.

\paragraph{Experimental Setup.} We use a virtual machine configured with Ubuntu 22.04, 4vCPUs, 64 GB of RAM, and 75 GB of disk space. This VM is provisioned on a university-maintained data center. The physical node it is hosted on has 2 10-core CPUs and 256 GB of RAM. We use the Linux \texttt{time} application to measure lifetime peak memory usage, and in-code timers to measure time for operations. As a part of the artifact, we provide a Docker image and a Dockerfile with Ubuntu 24.04 for the system configuration as a part of our artifact. All programs are compiled with either GCC 13.3.0 (in Ubuntu 24.04) or 11.4.0 (in Ubuntu 22.04). No optimization flags are passed to any C++ program. In the experiments for \texttt{Claim 1} and \texttt{Claim 2}, we take the arithmetic mean of 3 readings.

\subsection{Claim 1 and 2: Abstract Model Benchmarks}
\label{sec:abstract-model-benchmarks}

This section provides experimental results for \texttt{Claim 1} and \texttt{Claim 2}. We wrote 2 C++ classes that represent integer sets and points-to sets with integers, tively. There are two versions of each class, one with an underlying implementation in LHF, and the other a naive implementation with \texttt{std::set} and \texttt{std::map}\footnote{Using ordered containers allows us to utilize the STL set operation implementations and replicate what was present in VASCO-LFCPA.}. We have driver programs that use these classes, and accept file input containing instructions for either registering a set or performing union, intersection, or difference over sets. Second, we have Python programs that generate these instructions for each case.

The Python programs serve as the `ground source of truth' for the integer set operations, and were used to initially verify the output of the C++ programs. In the data collection phase, we disable all verification and debugging routines to obtain the cumulative time of each set operation and data creation/registration operation, and peak memory usage (PMU) of the lifetime of each run. While we don't consider comparing the Python generator and C++ programs to be an apples-to-apples comparison, we nonetheless include them because the Python script performs comparably in terms of operation time and memory.

\begin{figure}[t]
	\centering
	\includegraphics[width=0.95\textwidth]{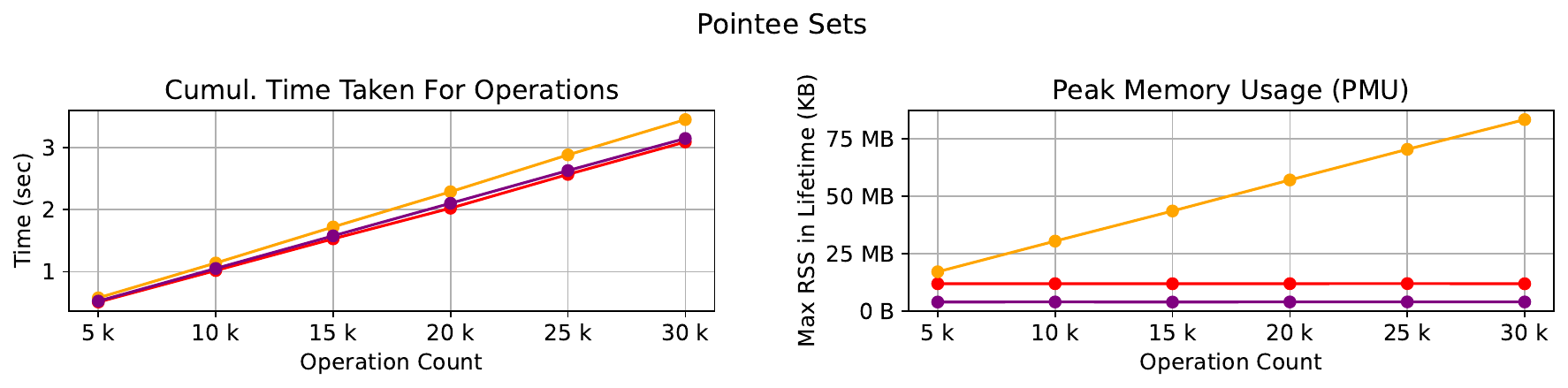}
	\includegraphics[width=0.95\textwidth]{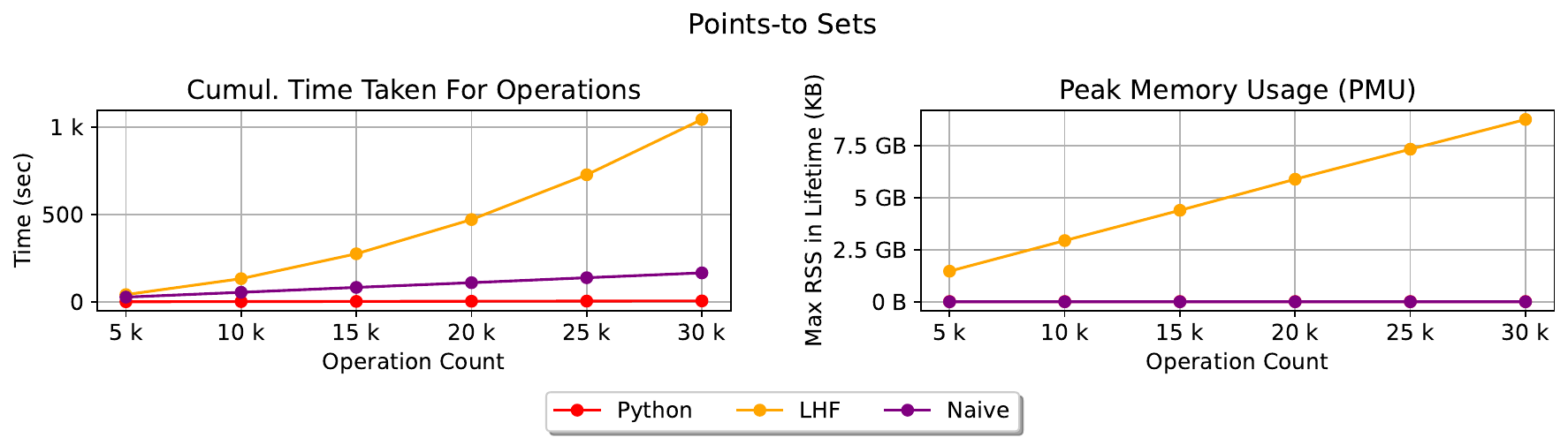}
	\caption{Results for \texttt{Claim 1}.}
	\label{table:claim1-evalmetrics}
\end{figure}

\paragraph{Claim 1.} We generate random sets of integers of random sizes, with a max integer value of 10,000 and a max size of 200 respectively. We pass them to the programs, and test them for up to 30,000 operations. As can be seen in~\autoref{table:claim1-evalmetrics}, LHF performs slightly worse in terms of time against the naive implementation when only pointee-sets are considered, but the memory usage of LHF drastically blows up due to the memoization and deduplication. When points-to sets are considered, due to the ad-hoc manner in which points-to sets are registered in LHF with one points-to pair inserted at a time, operation time blows up as well. Both perform worse against the Python generator script. Using LHF has little benefit in scenarios like this.

\begin{figure}[t]
	\centering
	\includegraphics[width=0.95\textwidth]{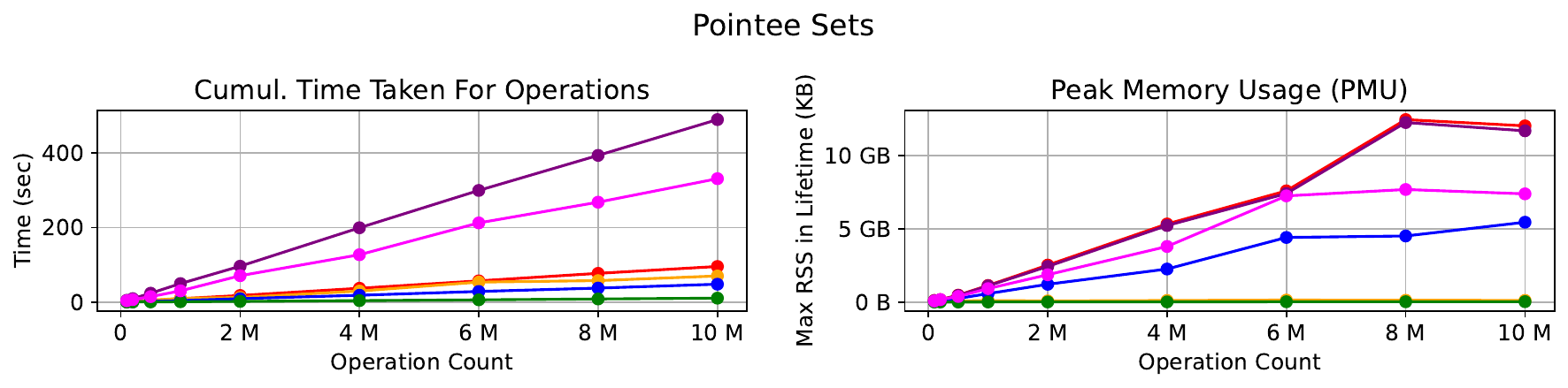}
	\includegraphics[width=0.95\textwidth]{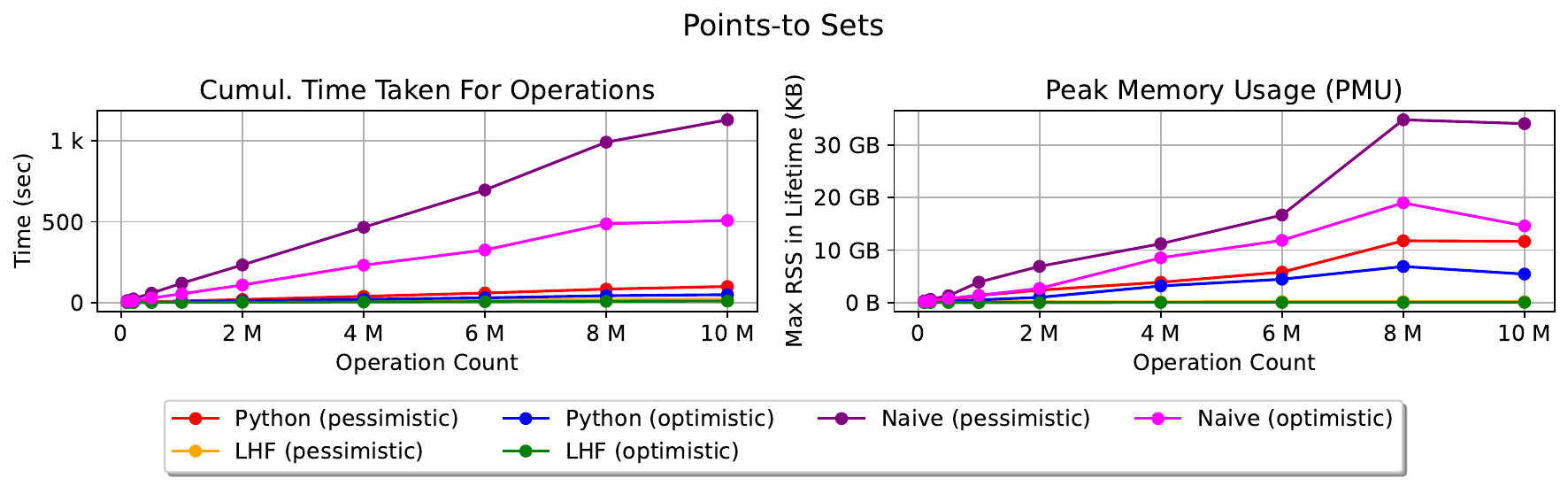}
	\caption{Results for \texttt{Claim 2}.}
	\label{table:claim2-evalmetrics}
\end{figure}

\begin{figure}[t]
	\centering
	\includegraphics[width=0.95\textwidth]{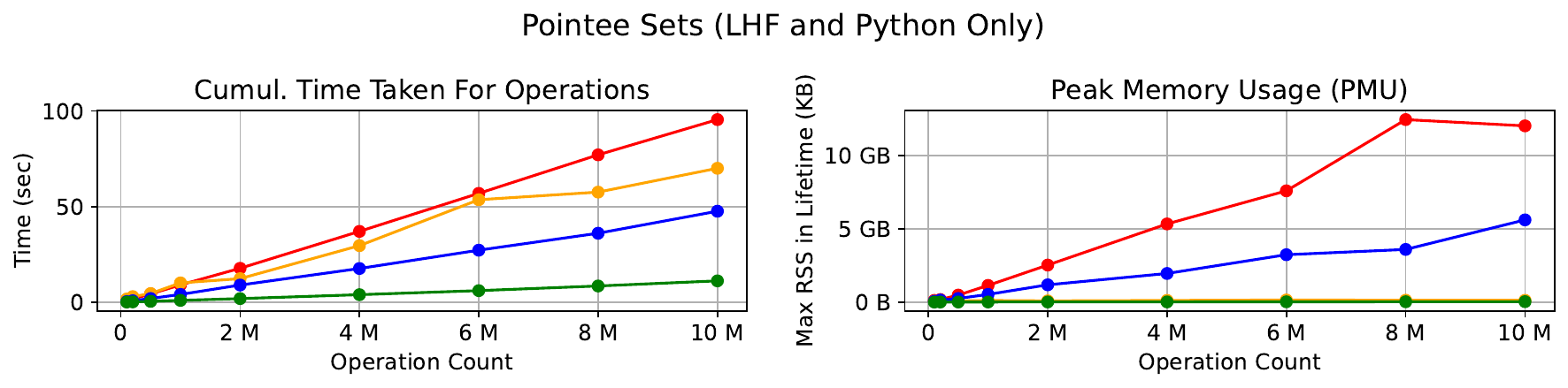}
	\includegraphics[width=0.95\textwidth]{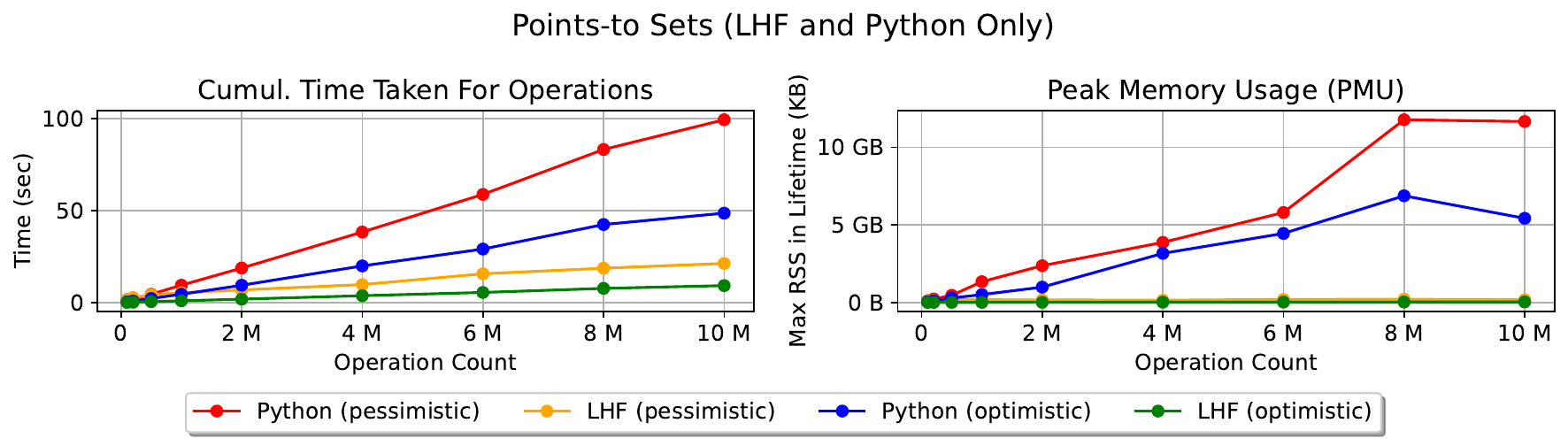}
	\caption{Results for \texttt{Claim 2} for only LHF and Python.}
	\label{table:claim2.5-evalmetrics}
\end{figure}

\paragraph{Claim 2: Most Pessimistic Case.} We generate a random number of sets as an initial corpus array of size 300, with the other constraints remaining the same as in the previous case. Then, we generate random operations by specifying indices within this corpus of data. We specify another instruction to randomly increase the corpus size by duplicating a random segment of data within the corpus, which the script replicates. The C++ programs are modified to store this initial data in a vector. The size of the vector is increased when specified by an instruction. We test all three implementations for up to 10 million operations. The results are illustrated in~\autoref{table:claim2-evalmetrics} and~\autoref{table:claim2.5-evalmetrics}, suffixed with `pessimistic' in the legend. As can be seen, LHF performs better than the naive version but not better than Python. However, memory consumed by LHF is a fraction of what the naive and the Python versions consume, as the number of operations increases. LHF had a PMU of \SI{125.3}{\mega\byte} after 10 million operations in the pointee set case and \SI{194}{\mega\byte} in the points-to set case. Speedup against Python ranges from 0.5x to 1.4x for pointee sets 0.5x to 4.7x for points-to sets as we approach 10M operations starting from 100,000.

We consider this to be the most pessimistic case because there is no redundancy in either the operations or the initial data. We believe LHF nonetheless manages to cope with it memory-wise asymptotically because redundancy in operations eventually exposes itself since the original data is finite. A point to note is that in the case of points-to sets, LHF wins out against Python by a large margin, which we believe might be because of the multilevel memoization.

\paragraph{Claim 2: Comparatively Optimistic Case.} We modify the Python script to generate a set $D$ of random values, one-fourth the corpus size, under the same constraints as before. The corpus array is initialized with empty sets, and values from $D$ are randomly assigned to the corpus' indices for three-fourths of the corpus size.

Next, we build a \textit{corpus of operations} by randomly pairing two corpus indices with a random operation. Unlike the data corpus, operations are always re-sampled rather than copied, and their growth is limited to less than ore equal to than the data increment. The script then samples instructions from this operation corpus.

These changes create a \textit{more optimistic case} for LHF, as these heuristics now guide the generation of redundant data and operations. Results in~\autoref{table:claim2-evalmetrics} and~\autoref{table:claim2.5-evalmetrics} show LHF achieving significantly better time performance than both Python and the naive implementation. The time improvement over Python is not constant as operations increase, suggesting possible data `saturation'. LHF had a PMU of \SI{32.1}{\mega\byte} after 10 million operations  in the pointee set case and \SI{49}{\mega\byte} in the points-to set case. Speedup against Python ranges from 3x to 4.2x for pointee sets 3.4x to 5.2x for points-to sets as we approach 10M operations starting from 100,000.

\begin{table}[t]
	\centering
	\caption{Results for \texttt{Claim 3}. All values are approximations.}
	\label{table:lfcpa-evalmetrics}
	\renewcommand{\arraystretch}{1}
	\begin{tabular}{l r r r r r}
		\toprule
		\thead{Benchmark} &
		\thead{Operation \\ Count} &
		\thead{Time \\ (Naive)} &
		\thead{Time \\ (LHF)} &
		\thead{PMU \\ (Naive)} &
		\thead{PMU \\ (LHF)} \\
		\midrule
			\texttt{mcf}        & \num{20485}          & \(\approx\)\SI{34}{\milli\second}   & \(\approx\)\SI{34}{\milli\second}   & \SI{59}{\mega\byte}   & \SI{58.1}{\mega\byte} \\
			\texttt{lbm}        & \num{117739}         & \SI{223}{\milli\second}             & \SI{197}{\milli\second}             & \SI{77.9}{\mega\byte} & \SI{67.5}{\mega\byte} \\
			\texttt{libquantum} & \(\approx\)\num{1.25e6}   & \SI{5839}{\milli\second}            & \SI{3206}{\milli\second}            & \SI{161.7}{\mega\byte} & \SI{98.8}{\mega\byte} \\
			\texttt{bzip2}      & \(\approx\)\num{71e6}     & \SI{50.6}{\minute}                  & \SI{22.9}{\minute}                  & \SI{12.3}{\giga\byte}  & \SI{794.2}{\mega\byte} \\
			\texttt{sjeng}      & \(\approx\)\num{310e6}    & \SI{8}{\hour}                       & \SI{3.4}{\hour}                     & \SI{47.3}{\giga\byte}  & \SI{2.6}{\giga\byte}  \\
			\texttt{hmmer}      & \(\approx\)\num{411.8e6}  & \SI{12.6}{\hour}                    & \SI{1.5}{\hour}                     & \SI{27.4}{\giga\byte}  & \SI{2.6}{\giga\byte}  \\
		\bottomrule
	\end{tabular}
	\renewcommand{\arraystretch}{1}
\end{table}

\subsection{Claim 3: Benchmarking VASCO-LFCPA}
\label{sec:pta}

We use the existing C++ implementation of VASCO-LFCPA as mentioned in~\autoref{sec:initial-approaches}, and modify it to use LHF. VASCO-LFCPA itself is implemented under LLVM 14 and uses another in-house helper library called SLIM\footnote{``Simplified LLVM IR Modelling''}. This is still in progress, and currently known to work only for simple C programs and the SPEC CPU 2006 benchmarks, requiring non-trivial modifications to LLVM IR files\footnote{Global variables need to be inserted as dummy assignments in the entry function to be processed.}. Although SPEC CPU 2006 is obsolete, this is irrelevant since our focus is on analysis of the program rather than execution. The results are neither sound nor deterministic, but we accept this as our goal is to evaluate LHF’s performance in a realistic setting. Given the difficulty in reproducibility, and the large code base (\(\approx\)20K lines) making the integration not as complete as we would like it to be, we present our findings primarily as an observation from this integration effort.

We execute VASCO-LFCPA on the benchmarks \texttt{mcf, lbm, libquantum, bzip2, sjeng,} and \texttt{hmmer} and record the time and peak memory usage. Due to the large amount of time it takes for each analysis round, we only took one reading for \texttt{sjeng} and \texttt{hmmer}. The results are shown in~\autoref{table:lfcpa-evalmetrics}. LFCPA outperforms the pre-existing naive implementation both in terms of time and memory. There is no clear, consistent trend for gains here, however.

\paragraph{Other Evaluations and Possible Observations.}

We provide a few other notes on evaluation in~\autoref{appendix:additional-metrics}. There are many observations that can possibly be made for LHF and should be actively explored, but for the sake of brevity and lack of mature work, we do not add them as part of our evaluations.

\section{Possible Applications of LHF}
\label{sec:applications}

We describe the applicability of LHF in more program analyses other than pointer analysis in~\autoref{appendix:similar-program-analysis}. Besides program analysis, the abstract model detailed in~\autoref{subsec:abstract-model:problem} is very similar to other topics in compiler design, such as parser generation and parsing. However, possibilities of problems similar or adjacent to this model have not yet been thoroughly explored. Nonetheless. we still present some ideas based on informal reasoning.

\paragraph{JSON Document Recursive Schema Generation.}
\label{subsec:applications:jsondoc}
JSONSchema~\cite{web-jsonschema} is a standard way to represent the structure of a JSON document. JSONSchemas can be defined in a recursive fashion much similarly in the way to a context-free grammar. This process can be automated on an example document by using LHF as an engine to memoize the types of the members of an object or array with a depth-first search, and then unique IDs can be assigned to structures that will denote their types. We describe a similar system for LHF constructions in~\autoref{appendix:cfgsignificance}.

\paragraph{Formal Logic Systems.}
\label{subsec:applications:reasoning-graph}
Problems found in compiler design have a lot of overlap with those that are pursued in formal logic. For example, e-graphs, as we saw in~\autoref{sec:applications}, are an essential part of satisfiability solvers like Z3~\cite{wiki-z3} and this can be reimplemented with LHF. The effectiveness of LHF in a scenario like this is still to be seen.

\section{Conclusions and Future Work}
\label{sec:conclusion}
We have presented and described \textit{LatticeHashForest} (LHF), a data structure meant for fast and efficient element-level operations on and storage of aggregate data. We have shown that the key difference between LHF and other similar data structures is that it allows for efficient element-level manipulation and retrieval of data on top of just having efficient operations on aggregate structures.

We have made available a full-featured C++ implementation of LHF, as well as several instances of evaluation on it. Most prominently, we have shown  speedups beyond 4x for inputs approaching 10 million, and a memory consumption reduction to a nearly negligible fraction of that in comparable implementations. We have also attempted to verbalize the problem that LHF solves in the hopes that similar problems from different domains can also be identified and solved using LHF or a data structure like LHF.

Our main goal with LHF is to create a solution that allows us to perform whole-program analysis efficiently. LHF aims to make it so that the availability of primary memory is no longer a constraining factor of the analysis. While LHF has already shown us substantial gains towards achieving these goals, further improvements and the implementation of features such as those in~\autoref{sec:other-features} and~\autoref{appendix:proposed-features}, we will likely be able to further enhance our solution, minimizing these constraints even more. In the future, we hope to find more applications of LHF beyond compiler design and optimization, and consequently, find more parallels of the core problem that LHF solves. 

\acks
\label{acks}
We would like to thank Aditi Raste, PhD scholar working with Prof. Uday Khedker, for her work on the C++ implementation of VASCO-LFCPA, which we have adapted for our use.

\appendix

\section{More Details on Pointer Analysis}
\label{appendix:pta}

This section is dedicated to providing a more detailed description of pointer analysis and attempting to describe why a precise and scalable pointer analysis is useful.

\subsection{Pointer Analysis, Points-to Analysis, and Alias Analysis}
\label{appendix:pta:naming}

Pointer analysis can be referred to by different names. One of the common names for it is \textit{alias analysis}~\cite{ptatutorial}. While the two terms may be used interchangeably, the key difference in alias analysis is that it determines whether two pointers alias to the same memory location, that is, generate equivalence classes of pointers, rather than directly answering where a variable may point to - neither a symmetric nor a reflexive relationship. Another important distinction is that in the former case, there is no explicit need to represent heap memory locations, whereas in the latter case, we do have to. A name that is synonymous with pointer analysis is `points-to analysis', and is used interchangeably with it in this document.

\subsection{Steensgaard's and Andersen's Analyses}

Steensgaard's and Andersen's analyses~\cite{steensgaard1996points, andersen1994program} are one of the oldest widely used pointer analysis algorithms. They compute points-to data by performing heavy summarizations of the data presented in all of the points in the control flow of a program at once. This typically results in very imprecise information. That is, it is not flow-sensitive.

The analysis we use, in contrast, is a flow-sensitive and context-sensitive analysis, which gives us a more precise result but at higher costs.

\subsection{Brief Description of Pointer Assignments}

\begin{figure}[t]
	\centering
	\includegraphics[width=\textwidth]{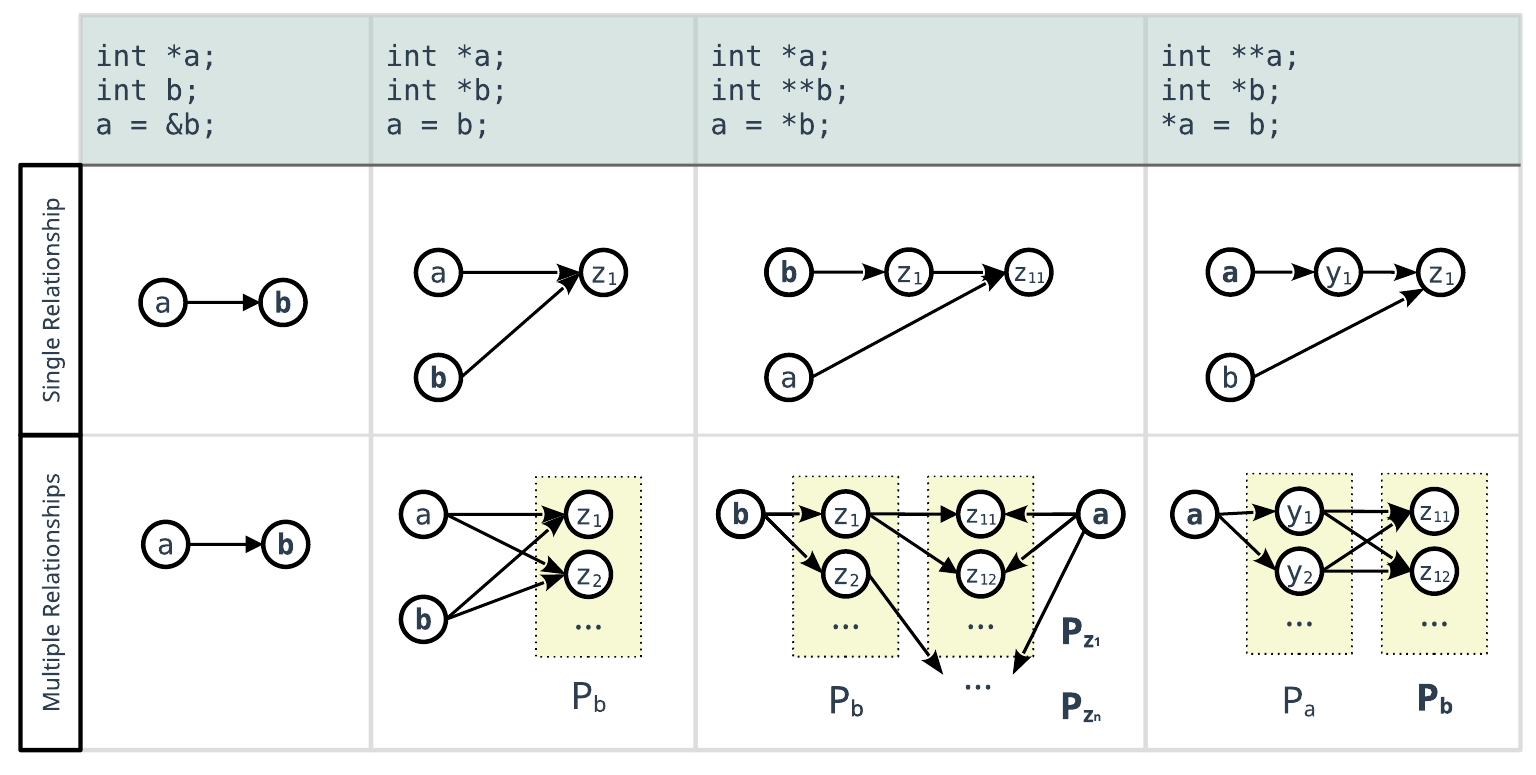}
	\caption{Examples of transitive propagation of relationships between pointers.}
	\label{fig:ptarel}
\end{figure}

Consider~\autoref{fig:ptarel}, which depicts a list of possible pointer assignments in the C language and the consequences on the points-to relationships. It is possible to see that the transitive relationships present between the pointers would cause the information in the analysis to increase non-linearly as it progresses.

One more point to note is that the pointer assignments cause updates at the \textit{element level}, if we are using a points-to set representation. Say there is a points-to set as follows:
\[\{
	a \rightarrow \{ q_1 \},
	b \rightarrow \{ q_2, q_3 \}
\}\]
At some point in the control flow, we assign $q_4$ to $b$. The points-to information thereafter becomes:
\[\{
	a \rightarrow \{ q_1 \},
	b \rightarrow \{ q_4 \}
\}\]
This update in points-to information is not an aggregate operation like union or intersection, but rather an element-level update on a single pointer $b$. Such operations are cumbersome to model with aggregate operations or with a set of points-to pairs as the representation.

Precise analysis information, paradoxically, can be less than the size of an imprecise analysis, but the difficulty incurred in computing precise information, and having to actually represent this precise information, is the reason why precise pointer analysis is not considered scalable. Our efforts are focused towards managing this computation and data overhead, and LHF is one part of managing those efforts.

For a more formal definition of the algorithm itself, please refer to~\autocite{lfcpa} and~\autocite{lfcpa-gen}.

\subsection{Soundness, Soundines and Scalability}

\textit{Soundness}, in the context of program analysis, is the property of an analysis result representing all possible program paths. Finding a scalable, precise points-to analysis that is also sound has been an unsolved problem, and is one of the problems LHF is trying to solve. 

The term \textit{Soundiness}, on the other hand, is used when an analysis is sound given some conditions. In recent years~\cite{soundiness}, it has been the trend to create partially sound analyses to analyze programs. An unsound analysis cannot be used for program transformation and optimization, but still may be useful for detecting vulnerabilities, limited code refactoring, and other topics.

Having a precise, sound, and scalable pointer analysis also opens the door towards efficient and precise whole-program analysis, because a precise analysis would contain the minimal possible relationships for function pointers rather than a large, imprecise summarization, reducing overhead for further interprocedural analysis.

\section{The Deduplication Mechanism}
\label{appendix:deduplication}

The deduplication mechanism, as described in~\autoref{subsec:initial-approaches:dedup}, is one of the main components of LHF.  The main purpose of the deduplicator is to assign a unique integer to each unique data item given to it. It has two main components: a storage vector and a hash table performing the mapping between items and IDs.

\paragraph{Structure of the Mapping Table.} For it to uniquely identify a data item, the deduplicator mandates 2 routines from the user:

\begin{description}
\item[Equality Comparator:] For simple data types, this may be simple in nature and may not even require explicit specification. However, in large or aggregate data structures like the property sets in LHF, such comparators have to be explicitly implemented and specified. This also implies that this is not necessarily a constant-time operation.
\item[Hash Function:] The hash table requires this for $O(1)$ lookup of values. Like in the case of the equality comparator, this is also not necessarily an $O(1)$ operation. 
\end{description}

This also surfaces another problem: the storage of keys themselves. The data items, as alluded to, can get large in size, and storing them as the keys of the item-to-integer mapping may result in very slow behavior if the hash table is reallocated or moved around, or the key-value pairs themselves are relocated.

Another problem lies in the case of performing a lookup. If the hash table implementation passes copies of the lookup parameter to the hashing function or any other internal component, this will also increase the amount of memory and time overhead of the deduplicator implementation. Hence, we keep pointers as our keys instead, making it a mapping from a pointer to an integer. The places we make these pointers point to are the storage locations in memory held by the storage vector when a new item is registered. The pointers, therefore, subsequently map to their respective identifier.

\paragraph{Structure of the Storage Vector.} The storage location of the actual data item is made to never change for the lifetime of the deduplicator object. This is handled by the storage vector, which is a resizable vector of C++ STL \texttt{std::unique\_ptr}s (smart pointers) that will free the held object upon destruction. The reason we don't simply just directly store the items as-is is because the semantics of the data item may not allow relocation, copying, or cause undefined behavior when the vector is resized while we are making the mapping table point to the elements of the vector. This makes the smart pointer approach sounder and easier to use.

\paragraph{Deduplication Flow Example.} Let's say the user wants to register some item A, and A is currently not in the deduplicator. The hashtable takes a pointer to A, tries to look it up, and it does not find it. The registration routine then proceeds to create a new \texttt{std::unique\_ptr} object, and copies or moves A into it depending on what the user prescribed as a parameter. The \texttt{std::unique\_ptr} object is then inserted into the storage array, and a new mapping is inserted with the storage location held by \texttt{std::unique\_ptr} as the parameter, and its index of insertion. Since this is a vector, the index acts as both an identifier and the location in the vector where the object can be found. Finally, the user is returned the index as the canonical identifier for A.

\paragraph{Usage in LHF.} While the deduplicator incurs an O(n) cost of hashing and equality lookup, we mitigate this by exploiting the repetitiveness of the data throughout the aggressive memoization that we form on it. We count on the data being sparse and redundantly computed and compared to get our efficiency gains.

\lstset{
  language={C++}
}
\begin{lstlisting}[caption={Sample implementation.}]
#include <cassert>
#include <cstddef>
#include <functional>
#include <iostream>
#include <memory>
#include <unordered_map>
#include <vector>

// Typedefs for convenience

template<typename T>
using UniquePointer = std::unique_ptr<T>;

template<typename T>
using Vector = std::vector<T>;

template<
	typename K,
	typename V,
	typename Hash = std::hash<K>,
	typename Equal = std::equal_to<V>>
using Map = std::unordered_map<K, V, Hash, Equal>;

using Index = std::size_t;

/// The deduplicator class

template<
	typename ItemT,
	typename ItemHash = std::hash<ItemT>,
	typename ItemEqual = std::equal_to<ItemT>>
class DeduplicatorExample {

	Vector<UniquePointer<ItemT>> data;
	Map<ItemT*, Index, ItemHash, ItemEqual> mapping;

public:
	// This inserts values into the deduplicator
	Index register_value(ItemT &v) {
		auto cursor = mapping.find(&v);
		if (cursor == mapping.end()) {
			UniquePointer<ItemT> s(new ItemT(v));
			data.push_back(std::move(s));
			Index ret = data.size() - 1;
			mapping.insert({data.at(ret).get(), ret});
			return ret;
		} else {
			return cursor->second;
		}
	}

	Index register_value(ItemT &&v) {
		auto cursor = mapping.find(&v);
		if (cursor == mapping.end()) {
			UniquePointer<ItemT> s(new ItemT(std::move(v)));
			data.push_back(std::move(s));
			Index ret = data.size() - 1;
			mapping.insert({data.at(ret).get(), ret});
			return ret;
		} else {
			return cursor->second;
		}
	}

	// This retrieves references from the deduplicator.
	const ItemT &get_value(Index idx) const {
		assert(idx >= 0);
		assert(idx < data.size());

		return *data.at(idx);
	}
};

// This is an example of what we can insert into a deduplicator.
struct Rectangle {
	int length;
	int breadth;

	Rectangle(int length, int breadth):
		length(length), breadth(breadth) {}

	// std::cout printing function for convenience
	friend std::ostream& operator<<(std::ostream& os, const Rectangle& r) {
		os << "Rectangle(" << r.length << ", " << r.breadth << ")";
		return os;
	}
};

// A 'Functor' struct that implements hashing for the Rectangle class.
struct RectangleHash {
	std::size_t operator()(const Rectangle *a) const {
		return
			std::hash<int>()(a->length) ^
			(std::hash<int>()(a->breadth) << 1);
	}
};

// A 'Functor' struct that implements equality comparison for the Rectangle
// class. This is unnecessary however and can be implemented by implementing
// operator== in Rectangle. Then in DeduplicatorExample, RectangleEqual can
// be substituted with std::equal_to<Rectangle>.
struct RectangleEqual {
	bool operator()(const Rectangle *a, const Rectangle *b) const {
		return (a->length == b->length) && (a->breadth == b->breadth);
	}
};

int main() {
	DeduplicatorExample<Rectangle, RectangleHash, RectangleEqual> dedup;
	Index a = dedup.register_value(Rectangle(3, 5));
	Rectangle rect(3, 5);
	Index b = dedup.register_value(rect);
	Index c = dedup.register_value(Rectangle(4, 5));
	Index d = dedup.register_value(Rectangle(4, 5));
	Index e = dedup.register_value(Rectangle(5, 5));
	Index f = dedup.register_value(Rectangle(5, 6));
	std::cout
		<< "Index of a = " << a << ", Value = " << dedup.get_value(a) << std::endl
		<< "Index of b = " << b << ", Value = " << dedup.get_value(b) << std::endl
		<< "Index of c = " << c << ", Value = " << dedup.get_value(c) << std::endl
		<< "Index of d = " << d << ", Value = " << dedup.get_value(d) << std::endl
		<< "Index of e = " << e << ", Value = " << dedup.get_value(e) << std::endl
		<< "Index of f = " << f << ", Value = " << dedup.get_value(f) << std::endl;
	return 0;
}

/*
Output:

Index of a = 0, Value = Rectangle(3, 5)
Index of b = 0, Value = Rectangle(3, 5)
Index of c = 1, Value = Rectangle(4, 5)
Index of d = 1, Value = Rectangle(4, 5)
Index of e = 2, Value = Rectangle(5, 5)
Index of f = 3, Value = Rectangle(5, 6)
*/
\end{lstlisting}

\section{Automated LHF Construction and Significance of a Context-Free Grammar Modelling of LHFs}
\label{appendix:cfgsignificance}

\begin{figure}[t]
	\centering
	\includegraphics[width=0.8\textwidth]{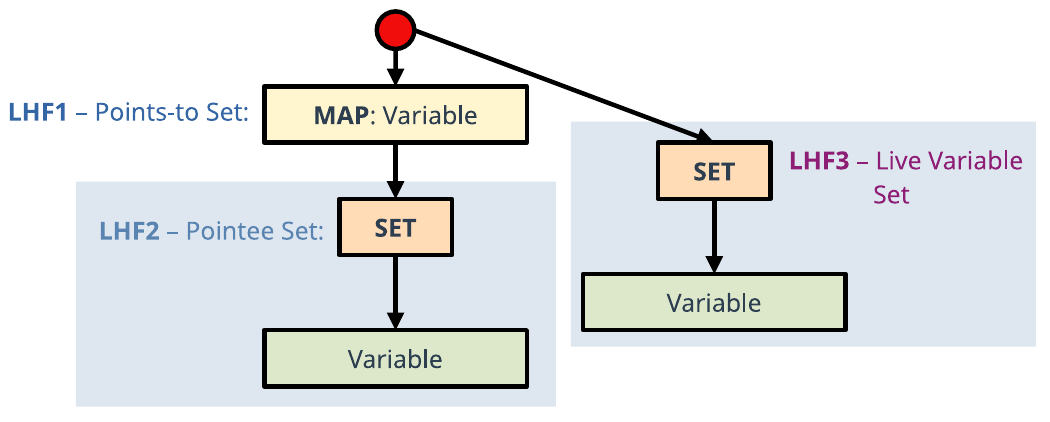}
	\caption{Illustration of the LHF construction found in LFCPA.}
	\label{fig:lfcpa-config}
\end{figure}

As mentioned earlier in~\autoref{sec:abstract-model} and~\autoref{sec:latticehashforest}, property sets can be nested in an unrestricted manner such that a key can point to multiple sets present in multiple LHFs. The ability to model this nesting as a context-free grammar allows us to perform optimizations on the construction of the LHF itself. \autoref{fig:lfcpa-config}~shows an illustration of the configuration of LHF for LFCPA that we mentioned in~\autoref{subsec:latticehashforest:lfcpa-nesting}. If we are able to express this configuration as a context-free grammar, it means that we will be able to use a program to programmatically determine the equivalence of the structures of \texttt{LHF3} and \texttt{LHF2} as shown in the diagram, simply by using a deduplicator-like mechanism and performing a bottom-up analysis of the supposed parse tree and generate the needed and optimized LHF declarations automatically.

We have implemented such an experimental C++ code generator tool\footnote{This can be found in the public GitHub repository at \url{https://github.com/aghorui/lhf/tree/master/src/builder_tool}, and in the artifact as \texttt{src/builder\_tool} in the LHF source code. The construction specifications as `blueprints' in the code.} in Python and is available as a part of the artifact. The implementation accepts JSON files as the specification and accepts abstract data structures, like maps and sets, to model the underlying specification. An example JSON specification is shown in~\autoref{appendix:cfgsignificance:sampledata}.

The implementation is, however, unstable and not all semantic rules that are possible in this grammar are covered. The headers for the C++ set operations in the abstract model benchmarks~(\autoref{sec:abstract-model-benchmarks}) were initially generated through this script. We hope for this script to be useful in quickly iterating over and testing out different constructions of LHF for the problem case.

\paragraph{Efficiency Gains from an LHF Construction.} The possibilities of modelling an LHF configuration in this manner, besides being able to consolidate similar sub-constructions, still need to be explored. Particularly, it still needs to be observed in what ways configurations can be made such that it exposes the \textit{`least amount of entropy'} at the highest level, like how we did with the points-to set representation. It also needs to be observed how higher and higher levels of indirection in a construction would affect performance.

\section{Additional Metrics for LHF}
\label{appendix:additional-metrics}

\begin{table}[t]
	\centering
	\renewcommand{\arraystretch}{1}
	\caption{List of LHF performance metrics.}
	\label{table:lhf-other-evalmetrics}
	\begin{tabular}{r  l}
		\toprule
		\thead{Metric} &
		\thead{Description} \\
		\midrule
		Hits &
		Operation already present in cache. \\
		Equal Hits &
		Operands are equal ($a \equiv b$).\\
		Subset Hits &
		One operand is a subset of another operand. \\
		Empty Hits &
		One of the operands is empty. \\
		\makecell[l]{Cold Misses\\ } &
		\makecell[l]{The result of the operation does not exist:\\ $c$ in $a \circ b = c$.} \\
		\makecell[l]{Edge Misses\\ } &
		\makecell[l]{The operation does not exist in cache, but the result exists:\\$a \circ b$ in $a \circ b = c$.} \\
		\bottomrule
	\end{tabular}
	\renewcommand{\arraystretch}{1}
\end{table}

Our evaluation in~\autoref{sec:evaluation} does not cover all the ways in which LHF can be evaluated. Here we specify some more metrics that can be extracted from LHF and what other behaviors can be tested in LHF.

\paragraph{Lattice Operation Hits and Misses} LHF may be evaluated much similarly in the way to a cache, in the sense that it also tries to look up data that may be already available to it. \autoref{table:lhf-other-evalmetrics}~lists the possible hits and misses that can be recorded for each operation in LHF in this context. This essentially is a profiling of how well is the incremental lattice or a set of relations is built by the LHF for the given data. This may also potentially allow us to profile the nature of the data that is being processed as input.

\paragraph{Data and Operation Saturation} One of the potentially useful observations to make for LHF in terms of the closed-world modeling is finding out if and when there is no longer any new data being created in the process, that is, when the data and operations saturate in the process. The results we show in our evaluation seem to suggest that this might be the case.

\paragraph{Data Size Distributions} It may also be useful to observe how small or large the property sets that are being operated on are. If the average property set is small, it may be possible that the problem LHF is being used for has high data entropy, or new data is being inserted into the system in a way LHF cannot cope with.

\paragraph{Data Access Patterns} Finding out Access patterns of data in an LHF would allow us to perform optimizations like clustering very frequently accessed data together, and save rarely accessed data to disk, saving memory. We elaborate on this idea in~\autoref{appendix:proposed-features:disk-backing}.

In summary, there are a lot of observations that can be made about LHF are observations that can be made about cache memory accesses or virtual memory page accesses as well. In the future, we hope to explore these areas of analysis to expand the scope of LHF and increase its efficiency.

\section{Proposed Features for LHF}
\label{appendix:proposed-features}

Besides the features we mention in~\autoref{sec:other-features}, we describe two proposed features below that we eventually plan to implement.

\subsection{Lazy-Insertion and Computations}
\label{subsec:other-features:lazy-insertion}

\begin{figure}[t]
	\centering
	\includegraphics[width=0.8\textwidth]{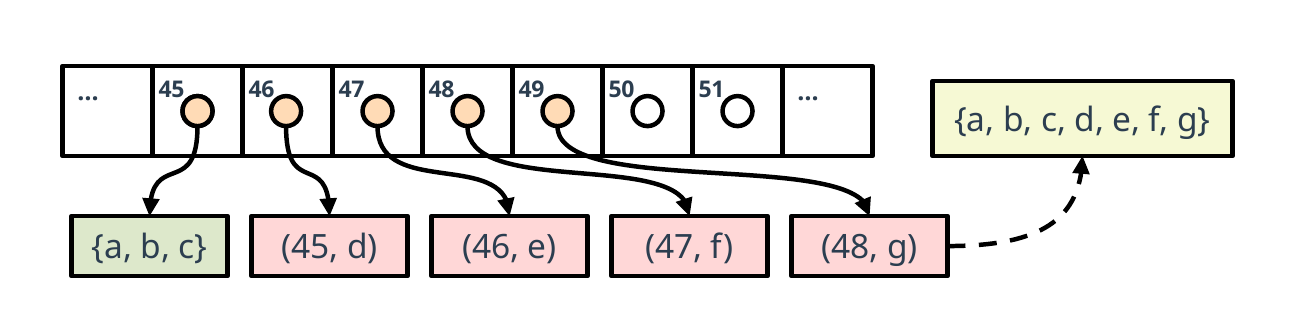}
	\caption{One possible lazy insertion mechanism.}
	\label{fig:lazy_insertions}
\end{figure}

This is similar in vein to a lot of the data structures we showed in~\autoref{sec:comparison}.

A pattern we observed in the case of integrating LHF into VASCO-LFCPA is that there were a lot of loops in place that would insert or remove a pointee from the set one at a time. This creates a lot of successive registrations of sets that may never be accessed again, which may cause bloat in memory.

If there are situations where such unitary insertions cannot be avoided, LHF may be augmented with a mechanism by which these successive incremental insertions would not be materially created, and will instead be lazily computed on access. Each of these potential computations is represented in the storage by a pointer/reference to the previous set and the element that is added or removed from the set, forming a `chain'. \autoref{fig:lazy_insertions} shows an illustration of one possible implementation of this mechanism.

This mechanism would possibly cause conflicts with the eviction mechanism described in~\autoref{sec:other-features}, as removal of sets would also cause the `lazy chains' to break, and therefore needs to be managed appropriately. The ways in which acyclic graph-like structures can be exploited to gain more efficiency in LHF are still to be seen.

\subsection{Disk-Backed Automatic Memory Page--Out}
\label{appendix:proposed-features:disk-backing}

\begin{figure}[t]
	\centering
	\includegraphics[width=0.8\textwidth]{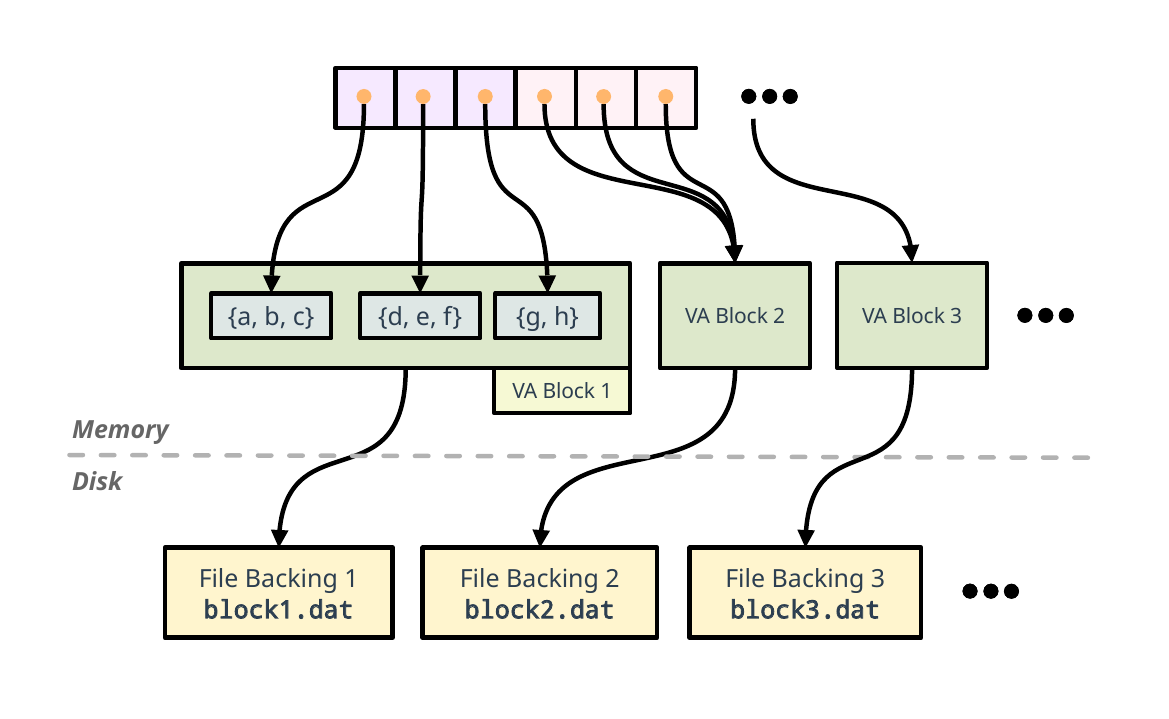}
	\caption{Disk-backed transparent memory management in LHF.}
	\label{fig:pageout}
\end{figure}

This is similar in vein to how paging works in operating systems, and serves to strengthen the virtual memory analogy that we use for LHF.

One of the main problems of performing points-to analysis on large programs is that memory quickly blows up as the analysis progresses, making it infeasible to perform. While LHF provides significant memory efficiency in this context, it may be possible that even this is not enough for sufficiently large programs.

One method we propose is to use blocks of \textit{memory-mapped regions  backed by files}~\cite{wiki-mmap} to store property sets in, and then based on speculation of the current operations being performed on the LHF, force them to be \textit{paged out} back to the file. When one of these regions is accessed, the file backing is brought back into memory. The way we plan to implement this in Linux is with the routine called \texttt{memadvise}~\cite{web-memadvise}, and particularly supply the advice value \texttt{MADV\_DONTNEED} to memory regions that have not been accessed in a long time (in an LRU fashion). This mechanism would have to be carefully managed such that we do not encounter thrashing when LHF is accessed frequently.

It still needs to be observed whether there is a requirement for such a mechanism for large programs, or if LHF by itself is sufficient enough with fairly large or moderate primary memory available.

\section{Brief LHF Implementation Details}
\label{appendix:lhf-implementation-details}

This section contains further clarification on the C++ implementation of LHF. We still consider LHF to be in the exploratory stages of work and may change drastically in the future as better methods, underlying data structures or algorithms are found for it. The details below reflect the design as it stood at the time of writing.

\subsection{The LHF Class}
\label{appendix:lhf-implementation-details:lhf-class}

The entire state of an LHF is stored as a single C++ class. All activities, such as registering new data, performing operations, etc. are implemented as member functions of this class. Specification of the type of data stored in the LHF is done through the template parameters of the class. The class declaration is as follows:

\lstset{
  language={C++}
}
\begin{lstlisting}
template <
	typename PropertyT,
	typename PropertyLess = DefaultLess<PropertyT>,
	typename PropertyHash = DefaultHash<PropertyT>,
	typename PropertyEqual = DefaultEqual<PropertyT>,
	typename PropertyPrinter = DefaultPrinter<PropertyT>,
	typename Nesting = NestingNone<PropertyT>>
class LatticeHashForest {
	...
\end{lstlisting}

The first parameter specifies the property type. The next three are used to specify how LHF should consider as its semantics for deduplicating the type and for set operations. The fifth parameter describes a pretty-printer for the property. Finally, \texttt{Nesting} specifies the nesting behavior of the LHF. In this case, the default value \texttt{NestingNone} specifies that the operation is not nested.

We recommend that LHF be declared as a static, global object, since it represents an domain of values, and the fact that it does not generally make sense to isolate the memoization that LHF does by making multiple instances of the same LHF with the same template parameters. This again leads into our virtual memory analogy, with it being a virtualization of values rather than a virtualization of memory locations.

\paragraph{Runtime Checks} LHF adds a series of assertions that can be used to detect violations of the data integrity or incorrect input. These can be disabled with a compile time switch, but the correctness of LHF in this case cannot be reasonably guaranteed.

\subsection{Sets as Sorted Vectors}
\label{appendix:lhf-implementation-details:sorted-vectors}

Sorted vectors offer a very straightforward way to perform various set operations with a worst-case complexity of $O(n)$, with algorithms similar to how the \textit{merge} operation is performed in the \textit{merge sort} algorithm. These algorithms can be performed with any ordered container and not just sorted vectors\footnote{Please refer to the implementations of C++ STL set operations, such as \texttt{std::set\_union}, \texttt{std::set\_intersection}, etc.}. We chose them, however, due to the following properties:

\begin{description}
\item[Immutability:] Because we enforce the constraint of immutability on property sets, it does not make much sense to choose a data structure with more overhead.
\item[Stable Representation and Predictability:] Sets implemented with tree data structures or hash tables may not have a canonical representation for their underlying data layout. Sorted vectors however, do. Data is completely sequential and adjacent, possibly allowing for predictable behavior, better cache locality and further memory optimizations.
\item[Random Access:] Sets can be accessed in any order needed, unrestricted by the container.
\item[No Restrictions on Element Data:] Values can be put into a sorted vector as long as they form a total order. No constraints besides having to be hashable, like having to map a binary encoding to each, are imposed on the values.
\item[Cheap Lookup:] While possibly not as cheap as lookup in a hash table, the sorted vector offers a key lookup complexity of $O(\log(n))$.
\end{description}

\paragraph{Registering Sets in LHF.} LHF expects the user to supply either a sorted vector or an ordered iterator to the data without any duplicates. The rationale behind this is that this allows the user more freedom to decide on peripheral data structures and algorithms for providing this sorted data. Nonetheless, we provide a removable runtime check for determining if the data does not meet these constraints.

\subsection{Implementation of Operations}
\label{appendix:lhf-implementation-details:operations}

Operations are implemented in the LHF class as members, as was mentioned earlier. Besides simply performing the operation, the member function is also responsible for performing all of the trivial checks, such as one of the sets being empty or two sets being equivalent.

This is likely not an optimal implementation or organization of this behavior, and may possibly be improved upon by implementing a `driver function' that would perform all of the common checks and memoization.

\subsection{Template Metaprogramming and Nesting}
\label{appendix:lhf-implementation-details:template}

The nesting behavior is implemented through a somewhat complex set of C++ template constructs. It is recommended to go through the source code to concretely understand its structure.

There are, by default, two values that can be specified as the \texttt{Nesting} parameter for an LHF, which are \texttt{NestingNone} and \texttt{NestingBase}. They implement the same semantics as \textit{NormalOperation} and \textit{NestedOperation} respectively as specified in~\autoref{sec:abstract-model:operations}. When \texttt{NestingBase} is specified, LHF is made to accept and store a tuple (\texttt{std::tuple}) of references to the LHFs specified by \texttt{NestingBase}, and each element of a property set additionally stores a tuple of child property set indices (i.e. the value) apart from the key. Please see~\autoref{appendix:sample-code} for examples.

To operate on two of such tuples of indices, we use an adaptation of the C++ STL function called \texttt{std::apply} to `unroll' them, and apply the specified operation on each corresponding pair of entries in each tuple to produce the resultant tuple of indices. To use the correct LHF with each corresponding pair of indices, we use a \textit{templated functor} that accepts a reference to an LHF, and the indices of the two sets. We let C++'s templating engine to perform all of this unpacking at compile-time.

Finally, to instruct an LHF whether to recurse or not, we keep a C++ \texttt{constexpr} boolean, that lets us, at compile time, disable or enable a branch within the code. We specify this boolean in \texttt{NestingNone} and \texttt{NestingBase}. It works similarly to specifying an \texttt{ifdef}, except that this is actually handled by the compiler and not the preprocessor. However, operation member function entries are annotated through the use of preprocessor macros to enable the nesting behavior.

Hence, the specifics of `looping' over each child property set index is determined at compile-time. We do not keep any runtime overhead for the nesting procedure itself. 

\subsection{Map Data Structure}
\label{appendix:lhf-implementation-details:maps}

LHF currently uses the C++ STL hash table (\texttt{std::unordered\_map}) implementation for implementing both the property set map as well as the operation maps. This implementation is known for not being very efficient. LHF merely mandates a particular interface to the map, and therefore can be swapped out for an alternate data structure by reimplementing the interface. 

We have done one such reimplementation with Intel TBB's~\cite{web-inteltbb} \texttt{concurrent\_hash\_map}. While this is in a working state, complete testing and performance comparisons with the existing implementation are still pending.

LHF currently does not have a requirement of an ordered map data structure.

\subsection{The Storage Vector}
\label{appendix:lhf-implementation-details:storage-vector}

LHF currently simply uses \texttt{std::vector} for its property set storage structure. This may result in bottlenecks in regards to reallocation if the vector gets too big and the reallocations become very frequent. We have not yet felt the need to implement an alternate mechanism yet, however.

Like in the case of maps, LHF also mandates a particular interface for the storage vector, and we have similarly created an alternate implementation with Intel TBB's \texttt{concurrent\_vector}. Its performance with respect to the original implementation still needs to be judged.

\subsection{Extending LHF}
\label{appendix:lhf-implementation-details:extending}

We expect users to extend the LHF class to implement new operations, functions that need access to the private API and so on. The heft of features and requirements do make the task of optimally implementing a new operation complicated, but we hope to make this easier as a part of our future work.

Our integration of LHF in VASCO-LFCPA has such examples of extensions to enable custom behaviors.

\section{Similar Program Analysis Use-Cases}
\label{appendix:similar-program-analysis}

A significant amount of work in our research group at IIT Bombay involves data structures and mechanisms that we consider to be potential good use cases for LHF. Although the integration work for LHF onto the existing implementations and subsequent testing is yet to be completed, we believe that the integration will lead to good performance and memory efficiency gains. The following examples serve to strengthen our case for LHF.

\subsection{Generalized Points-to Graphs (GPG)}
Generalized Points-to Graphs~\cite{gpg-analysis} represent points-to relationships with two integers denoting the amount of indirection between two variables instead of the concrete set of possible relationships. This summarizes the pointer assignment information within a given control flow graph, which then gets progressively simplified by an analysis to represent the points-to information as compactly as possible.

In the current implementation, the edges are represented as a tuple of 4 values. These tuples are put into a set to denote a graph, and these graphs are put into another set to represent the summaries for control-flow blocks. This set-of-set-of-sets structure isn't directly modelled by LHF, but is possible to model implicitly, that is, make the parent LHFs store a set of indices within the child LHFs. While there are no nesting methods used here directly, we are still able to represent aggregate nested data structures.

It may also be possible that we may change the representation structure entirely and use a representation similar to what we do for points-to sets. The different possible configurations have to be explored to see which exposes the maximum redundancy in data.

\subsection{Heap Reference Analysis (HRA)}
Heap Reference Analysis~\cite{hra} is a program analysis that determines possible access paths of the members of an object allocated in the heap. For example, consider a linked list node with two elements, \texttt{data} and \texttt{next}, where \texttt{next} points to another linked list node allocated somewhere. We may have a statement like \texttt{node = node.next} within a loop at some point in the program. HRA builds a \textit{deterministic finite automaton} (DFA) that represents this pattern. This means that we are passing around a set of DFAs that represent these access paths across the control flow of the program.

We consider this to be a good use case for LHF because, like in points-to analysis, we are simply propagating graphs across the program, and there is potential for leveraging the redundancies and sparseness of data in the analysis with LHF.

\section{Sample Source Code Listings}
\label{appendix:sample-code}

Source code examples referenced in the main text and elsewhere are placed over here.

\begin{lstlisting}[caption={Sample JSON for points-to data (with integers).}, label={appendix:cfgsignificance:sampledata}, language={} ]
{
	"lhf_version": "0.2.0",
	"namespace": "lfcpa",
	"scalar_definitions": {
		"integer": {
			"name": "long"
		}
	},
	"entity_definitions": {
		"liveset": {
			"name": "LiveSetStore",
			"interface": "LiveSet",
			"user_exposed": true,
			"operations": []
		},
		"pointstoset": {
			"name": "PointsToSetStore",
			"interface": "PointsToSet",
			"user_exposed": true,
			"operations": [
				// (This does not get used in the current implementation)
				{ "name": "update_pointees", "recursive": false }
			]
		},
		"pointeeset": {
			"name": "PointeeSetStore",
			"interface": "PointeeSet",
			"user_exposed": true,
			"operations": []
		}
	},
	"blueprint": [
		{
			"type": "set",
			"value": {
				"type": "scalar",
				"value": "integer"
			},
			"use_as": "liveset"
		},
		(*@\newpage@*)
		{
			"type": "map",
			"key": "integer",
			"value": {
				"type": "set",
				"value": {
					"type": "scalar",
					"value": "integer"
				},
				"use_as": "pointeeset"
			},
			"use_as": "pointstoset"
		}
	]
}
\end{lstlisting}

\lstset{
  language={C++}
}
\begin{lstlisting}[caption={LHF demonstration example code.}, label={appendix:sample-code:simple} ]
#include <iostream>
#include "lhf/lhf.hpp"

// Typedefs for convenience
using LHF = lhf::LatticeHashForest<int>;
using Index = LHF::Index;
using Elem = LHF::PropertyElement;

int main() {
	// LHF object
	LHF l;

	Index a = l.register_set({1, 2, 3});
	Index b = l.register_set({1, 2, 4});

	Index c = l.set_union(a, b);

	std::cout
		<< "Set {1, 2, 3} has index " << a << std::endl
		<< "Set {1, 2, 4} has index " << b << std::endl
		<< std::endl;

	std::cout << "Contents of Index " << c << ": ";

	for (const Elem &i : l.get_value(c)) {
		std::cout << i << " ";
	}

	std::cout << std::endl << std::endl;

	// This will dump all of the data present in the LHF.
	std::cout << l.dump();
}

/*
Output:

Set {1, 2, 3} has index 1
Set {1, 2, 4} has index 2

Contents of Index 3: 1 2 3 4

{
    Unions: (Count: 1)
      {(1,2) -> 3}

    Differences:(Count: 0)

    Intersections: (Count: 0)

    Subsets: (Count: 2)
      (2,3) -> sub
      (1,3) -> sub

    PropertySets: (Count: 4)
      0 : { }
      1 : { 1 2 3 }
      2 : { 1 2 4 }
      3 : { 1 2 3 4 }
}
*/
\end{lstlisting}

\lstset{
  language={C++}
}
\begin{lstlisting}[caption={Nested LHF demonstration example code.}, label={appendix:sample-code:points-to}]
#include <iostream>
#include <ostream>

#include "lhf/lhf.hpp"

// Typedefs for convenience
using ChildLHF = lhf::LatticeHashForest<int>;
using ChildIndex = typename ChildLHF::Index;
using ChildElem = typename ChildLHF::PropertyElement;

using LHF =
	lhf::LatticeHashForest<
		int,
		lhf::DefaultLess<int>,
		lhf::DefaultHash<int>,
		lhf::DefaultEqual<int>,
		lhf::DefaultPrinter<int>,
		lhf::NestingBase<int, ChildLHF>>;

using Index = typename LHF::Index;
using Index = typename LHF::Index;
using Elem = typename LHF::PropertyElement;

int main() {

	ChildLHF cl;
	LHF l(LHF::RefList{cl});

	ChildIndex ca = cl.register_set({3, 4, 5});
	ChildIndex cb = cl.register_set({4, 5, 6});

	Index a = l.register_set({
		{2, {ca}},
		{3, {ca}},
		{4, {cb}}
	});

	Index b = l.register_set({
		{4, {ca}},
		{5, {ca}}
	});

	Index result = l.set_union(a, b);

	std::cout << "Set a index: " << a << std::endl
	          << "Set b index: " << b << std::endl
	          << "Result set index: " << result << std::endl
	          << "Result set contents: " << std::endl;

	for (const Elem &i: l.get_value(result)) {
		std::cout << "  " << i << " :: { ";
		// We are retrieving the Child LHF index from the value tuple, which
		// is why we need std::get.
		for (const ChildElem &j: cl.get_value(std::get<0>(i.get_value()))) {
			std::cout << j << " ";
		}
		std::cout << "}" << std::endl;
	}

	std::cout << std::endl;

	std::cout << "@@@@ Pointee LHF @@@@" << std::endl;
	std::cout << cl.dump() << std::endl << std::endl;
	std::cout << "@@@@ Points-to LHF @@@@" << std::endl;
	std::cout << l.dump() << std::endl;

	return 0;
}

/*
Output:

Set a index: 1
Set b index: 2
Result set index: 3
Result set contents:
  2 -> [ 1 ] :: { 3 4 5 }
  3 -> [ 1 ] :: { 3 4 5 }
  4 -> [ 3 ] :: { 3 4 5 6 }
  5 -> [ 1 ] :: { 3 4 5 }

@@@@ Pointee LHF @@@@
{
    Unions: (Count: 1)
      {(1,2) -> 3}

    Differences:(Count: 0)

    Intersections: (Count: 0)

    Subsets: (Count: 2)
      (2,3) -> sub
      (1,3) -> sub

    PropertySets: (Count: 4)
      0 : { }
      1 : { 3 4 5 }
      2 : { 4 5 6 }
      3 : { 3 4 5 6 }
}


@@@@ Points-to LHF @@@@
{
    Unions: (Count: 1)
      {(1,2) -> 3}

    Differences:(Count: 0)

    Intersections: (Count: 0)

    Subsets: (Count: 2)
      (2,3) -> sub
      (1,3) -> sub

    PropertySets: (Count: 4)
      0 : { }
      1 : { 2 -> [ 1 ] 3 -> [ 1 ] 4 -> [ 2 ] }
      2 : { 4 -> [ 1 ] 5 -> [ 1 ] }
      3 : { 2 -> [ 1 ] 3 -> [ 1 ] 4 -> [ 3 ] 5 -> [ 1 ] }
}
*/
\end{lstlisting}

\lstset{
  language={C++}
}
\begin{lstlisting}[caption={Nested LHF demonstration with two children.},label={appendix:sample-code:points-to-2elem}]
#include <iostream>
#include <ostream>

#include "lhf/lhf.hpp"

// Typedefs for convenience
using Child1LHF =
	lhf::LatticeHashForest<double>;

using Child1Index = typename Child1LHF::Index;
using Child1Elem = typename Child1LHF::PropertyElement;

using Child2LHF =
	lhf::LatticeHashForest<int>;

using Child2Index = typename Child2LHF::Index;
using Child2Elem = typename Child2LHF::PropertyElement;


using LHF =
		lhf::LatticeHashForest<
			int,
			lhf::DefaultLess<int>,
			lhf::DefaultHash<int>,
			lhf::DefaultEqual<int>,
			lhf::DefaultPrinter<int>,
			lhf::NestingBase<int, Child1LHF, Child2LHF>>;

using Index = typename LHF::Index;
using Elem = typename LHF::PropertyElement;

int main() {
	Child1LHF cl1;
	Child2LHF cl2;
	LHF l(LHF::RefList{cl1, cl2});

	Child1Index a1 = cl1.register_set_single(212.23);
	Child1Index b1 = cl1.register_set({22.2, 33.122, 33.4});
	Child1Index g1 = cl1.register_set({34.121});

	Child2Index a2 = cl2.register_set_single(212);
	Child2Index b2 = cl2.register_set({22, 33, 34});
	Child2Index g2 = cl2.register_set({34});

	Index c = l.register_set({
		{12, {a1, a2}},
		{14, {a1, b2}},
	});

	Index f = l.register_set_single({12, {b1, b2}});
	Index e = l.register_set_single({12, {g1, g2}});

	Index result = l.set_union(c, f);
	std::cout << "Result set index: " << result << std::endl
	          << "Result set contents: {" << std::endl;

	for (const Elem &i: l.get_value(result)) {
		std::cout << "    " << i << " :: { " << std::endl
		          << "        Child 1: ";
		// We are retrieving the Child LHF index from the value tuple, which
		// is why we need std::get.
		for (const Child1Elem &j: cl1.get_value(std::get<0>(i.get_value()))) {
			std::cout << j << " ";
		}
		std::cout << std::endl
		          << "        Child 2: ";
		for (const Child2Elem &j: cl2.get_value(std::get<1>(i.get_value()))) {
			std::cout << j << " ";
		}
		std::cout << std::endl;
		std::cout << "    }" << std::endl;
	}

	std::cout << "}" << std::endl;

	std::cout << "@@@@ Child 1 LHF @@@@" << std::endl;
	std::cout << cl1.dump() << std::endl << std::endl;
	std::cout << "@@@@ Child 2 LHF @@@@" << std::endl;
	std::cout << cl2.dump() << std::endl << std::endl;
	std::cout << "@@@@ Main LHF @@@@" << std::endl;
	std::cout << l.dump() << std::endl;

	return 0;
}

/*
Output:

Result set index: 4
Result set contents: {
    12 -> [ 4 4 ] :: {
        Child 1: 22.2 33.122 33.4 212.23
        Child 2: 22 33 34 212
    }
    14 -> [ 1 2 ] :: {
        Child 1: 212.23
        Child 2: 22 33 34
    }
}
@@@@ Child 1 LHF @@@@
{
    Unions: (Count: 1)
      {(1,2) -> 4}

    Differences:(Count: 0)

    Intersections: (Count: 0)

    Subsets: (Count: 2)
      (2,4) -> sub
      (1,4) -> sub

    PropertySets: (Count: 5)
      0 : { }
      1 : { 212.23 }
      2 : { 22.2 33.122 33.4 }
      3 : { 34.121 }
      4 : { 22.2 33.122 33.4 212.23 }
}


@@@@ Child 2 LHF @@@@
{
    Unions: (Count: 1)
      {(1,2) -> 4}

    Differences:(Count: 0)

    Intersections: (Count: 0)

    Subsets: (Count: 2)
      (2,4) -> sub
      (1,4) -> sub

    PropertySets: (Count: 5)
      0 : { }
      1 : { 212 }
      2 : { 22 33 34 }
      3 : { 34 }
      4 : { 22 33 34 212 }
}


@@@@ Main LHF @@@@
{
    Unions: (Count: 1)
      {(1,2) -> 4}

    Differences:(Count: 0)

    Intersections: (Count: 0)

    Subsets: (Count: 2)
      (2,4) -> sub
      (1,4) -> sub

    PropertySets: (Count: 5)
      0 : { }
      1 : { 12 -> [ 1 1 ] 14 -> [ 1 2 ] }
      2 : { 12 -> [ 2 2 ] }
      3 : { 12 -> [ 3 3 ] }
      4 : { 12 -> [ 4 4 ] 14 -> [ 1 2 ] }
}
*/
\end{lstlisting}

\printbibliography

\end{document}